\documentclass[aps,prc,preprint,superscriptaddress,showpacs]{revtex4}
\begin{document}
\title{Role of five-quark components in radiative and strong decays of the $\Lambda(1405)$ resonance}

\author{C. S. An}
\email[]{chunsheng.an@cea.fr}
\affiliation{Institut de Recherche sur les
lois Fondamentales de l'Univers, DSM/Irfu, CEA/Saclay, F-91191 Gif-sur-Yvette, France}

\author{B. Saghai}
\email[]{bijan.saghai@cea.fr}
\affiliation{Institut de Recherche sur les
lois Fondamentales de l'Univers, DSM/Irfu, CEA/Saclay, F-91191 Gif-sur-Yvette, France}

\author{S. G. Yuan}
\email[]{yuanhadron@impcas.ac.cn}
\affiliation{Institute of Modern Physics, Chinese Academy of Sciences,
Lanzhou 730000, China}

\author{Jun He}
\email[]{junhe@impcas.ac.cn}
\affiliation{Institute of Modern Physics, Chinese Academy of Sciences,
Lanzhou 730000, China}

\date{\today}

\begin{abstract}

Within an extended chiral constituent quark model, three- and five-quark structure of the $S_{01}$
resonance $\Lambda(1405)$ is investigated. Helicity amplitudes for the electromagnetic decays
($\Lambda(1405) \to \Lambda(1116)\gamma$, $\Sigma(1194)\gamma$), and transition amplitudes for
strong decays ($\Lambda(1405)\to\Sigma(1194)\pi$, $ K^{-}p$) are derived, as well as the relevant
decay widths.
The experimental value for the strong decay width,
$\Gamma_{\Lambda(1405)\to (\Sigma \pi)^\circ}=50\pm 2$ MeV, is well reproduced with about 50\%
of five-quark admixture in the $\Lambda(1405)$.
Important effects due to the configuration mixings among $\Lambda^{2}_{1}P_{A}$,
$\Lambda^{2}_{8}P_{M}$ and $\Lambda^{4}_{8}P_{M}$ are found.
In addition, transitions between the three- and
five-quark components in the baryons turn out to be significant in
both  radiative and strong decays of the $\Lambda(1405)$ resonance.

\end{abstract}

\pacs{ 12.39.-x, 14.20.Jn, 13.30.Eg, 13.40.Hq}
\maketitle

\section{Introduction}
The structure and properties of the $S_{01}$ resonance
$\Lambda(1405)$, discovered in 1960's, is still one of the puzzling issues in hadron physics.
In the literature the $\Lambda (1405)$ is considered as an {\it s}-channel resonance \cite{DeGrand:1976kx}
or as a quasi-bound ($\bar{K}N,\Sigma \pi$) state
\cite{Dalitz:1960du,Dalitz:1967fp,Schnick:1987is,Zhong:1988gn,Siegel:1988rq,Tanaka:1992gj,Siegel:1994mb,Kimura:2000sm,Oset:2001cn,Wang:2006cfa,Jido:2009jf,Hyodo:2009np}.
In quark-model approaches, this hyperon is treated as a pure $|qqq\rangle$-state \cite{Jones:1977yk,Koniuk:1979vy,Darewych:1983yw,Darewych:1985dc,Kaxiras:1985zv,VanCauteren:2005sm,Yu:2006sc},
or still as an admixture of $|q^3$+$q^4$$\overline{q}\rangle$ configuration \cite{He:1993et,Arima:1994bv}.
Other approaches take this hyperon as an ``elementary'' field
\cite{Lee:1994jj} or as a quasi-bound state \cite{Kaiser:1995eg} using
chiral perturbation theory, or consider it as composed of an SU(2) soliton and
a kaon bound in an S-wave \cite{Schat:1994gm}.

In recent years, possible unconventional or exotic structure for that resonance has received
significant attention, suggesting the presence of states other than pure three-quark configuration.

QCD-sum Rules framework has been applied to
investigate \cite{Liu:1984dp,Choe:1997wz,Kondo:2006xz,Nakamura:2008zzc,Kisslinger:2009dr}
the nature of the $\Lambda$(1405). Using the $\Sigma^\circ \pi^\circ$ multiquark interpolation
field the mass of that resonance is overestimated by about 100 MeV \cite{Choe:1997wz}.
Introducing \cite{Kondo:2006xz} coupling between positive- and negative-parity baryons
within the flavor-octet hyperons leads to the conclusion that the $\Lambda$(1405) is not the
parity partner of the $\Lambda$ and may be a flavor-singlet or exotic state.
Mixing of three- and five-quark Fock components attributes \cite{Nakamura:2008zzc} to this
latter 90\% of occupations, employing a non-unique flavor-singlet operator for it,
composed of two flavor diquarks and one antiquark.
Moreover, a recent work \cite{Kisslinger:2009dr} predicts that resonance as an exotic $[udsg]$
strange hybrid and the mass of the lowest strange hybrid with $IJ^P = 0(1/2)^-$ turns out to
be 1407 MeV.

Various lattice QCD calculations  \cite{Melnitchouk:2002eg,Nemoto:2003ft,Lee:2002gn,Burch:2006cc,Ishii:2007ym,Takahashi:2009ik}
have been devoted to predict the mass of $\Lambda$(1405) and come up with masses higher than
the observed one by 300-400 MeV. An interesting outcome of those works is nevertheless the need
for five-quark components  in $\Lambda$(1405).

Investigations of the radiative and strong decay processes of baryons offer an appropriate case study
in getting reliable insights to their internal structure.
Several authors have studied the decay properties of the $\Lambda(1405)$ within constituent quark models
\cite{Koniuk:1979vy,Darewych:1983yw,Darewych:1985dc,Kaxiras:1985zv,VanCauteren:2005sm,Yu:2006sc}.
However, the calculated  strong decay width of the $\Lambda(1405)$ in the
traditional constituent quark model turns out to be much smaller than the
value $\Gamma=50 \pm 2$ MeV reported by the Particle Data Group (PDG) \cite{Amsler:2008zzb}.

Recently, extended constituent quark models, which include higher Fock components, have been developed to
describe the properties of baryon resonances \cite{Li:2005jn,Li:2005jb,Li:2006nm,JuliaDiaz:2006av,An:2008xk,An:2009uv}.
Those approaches strongly support the existence of significant genuine non-perturbative five-quark
components in baryons (for a recent concise review see Ref. \cite{Zou:2010tc}) and
provide much better descriptions for the electromagnetic and strong decays of
$\Delta(1232)$ \cite{Li:2005jn,Li:2005jb}, $N(1440)$ \cite{Li:2006nm,JuliaDiaz:2006av} and
$N(1535)$ \cite{An:2008xk,An:2009uv}.

Here we investigate the relevance of five-quark components in $\Lambda(1405)$, within
an extended chiral constituent quark approach.
The orbital-flavor-spin configuration for the four-quark
subsystem of the five-quark components in $\Lambda(1405)$, with lowest energy
being $[31]_{XFS}[4]_{X}[211]_{F}[22]_{S}$ \cite{An:2008tz,Helminen:2000jb},
allows for $u\bar{u}$, $d\bar{d}$ and $s\bar{s}$ components in this resonance,
while the lowest energy five-quark component in the $S_{11}$ nucleon resonance
$N(1535)$ can only be the $s\bar{s}$ component \cite{An:2008xk,An:2008tz}.
Those features shed a light on the observed mass ordering of $\Lambda(1405)$ and
$N(1535)$, which cannot be described wihtin conventional constituent quark models.

In this work we focus on the
radiative and strong decays widths of the $\Lambda(1405)$ in a truncated Fock space,
which includes three- and five-quark components, as well as configuration mixings among them,
namely, $qqq \leftrightarrow qqqq \bar{q}$ transitions (here, we have omitted the $\gamma^*$ or the meson,
$\pi$ and $K$, which intervene in those transitions).
We find that the mixing mechanism contributes significantly to both strong and radiative decays.

The manuscript is organized in the following way. The wave
functions for the three- and five-quark components in
$\Lambda(1405)$ and that in the $SU(3)$ octet baryons are given in
Section \ref{wf}. In Section \ref{fmm}, we give a brief account of
to the formalism for the radiative and strong decays in the extended
chiral constituent quark model. The numerical results are presented and discussed
in Section \ref{nr}. Finally, Section \ref{fn} contains our
conclusions.

\section{Wave function model}
\label{wf}

In our extended chiral constituent quark model, we assume that the
wave function for a baryon can be expressed as
\begin{eqnarray}
|B\rangle&=&A_{(B)3q}|qqq\rangle +A_{(B)5q}\sum_{i}A_{i}|qqqq_{i}
\bar{q_{i}}\rangle+\cdot\cdot\cdot\, .
\label{wfb}
\end{eqnarray}
Here $A_{(B)3q}$ and $A_{(B)5q}$ are the amplitudes for the 3-quark
and 5-quark components, respectively, in the corresponding baryon.
If we neglect higher Fock components, then
$A_{(B)3q}^2+A_{(B)5q}^2=1$. The sum over $i$ runs over all the
possible $qqqq_{i}\bar{q_{i}}$ components ($i=u,d,s$), and the
factors $A_{i}$ denote the coefficient for the corresponding
 $qqqq_{i}\bar{q}_{i}$ component, implying $\sum_{i}A_{i}^{2}=1$.

In this paper, we consider the $S_{01}$ resonance $\Lambda(1405)$ to
be an admixture of the configurations $\Lambda_{1}^{2}P_{A}$,
$\Lambda_{8}^{2}P_{M}$. and $\Lambda_{8}^{4}P_{M}$.
We also assume the $SU(3)$ octet baryons to be an admixture of
$B^{2}_{8}S_{S}$, $B^{2}_{8}S'_{S}$, and $B^{2}_{8}S_{M}$ configurations.
Concerning the mixing probability amplitudes for these latter configurations,
we employ, for simplicity, the ones proposed in Refs.~\cite{Koniuk:1979vy,Isgur:1978xj}
\begin{eqnarray}
|\Lambda(1405)\rangle&=&0.90|\Lambda^{2}_{1}P_{A}\rangle-0.43|\Lambda^{2}_{8}P_{M}\rangle+0.06|\Lambda_{8}^{4}P_{M}\rangle\,,
\label{cmc1}\\
|\Lambda(1116)\rangle&=&0.93|\Lambda^{2}_{8}S_{S}\rangle+0.30|\Lambda^{2}_{8}S'_{s}\rangle-0.20|\Lambda_{8}^{2}S_{M}\rangle\,,
\label{cmc2}\\
|\Sigma(1193)\rangle&=&0.95|\Sigma^{2}_{8}S_{S}\rangle+0.18|\Sigma^{2}_{8}S'_{S}\rangle-0.16|\Sigma_{8}^{2}S_{M}\rangle\,,
\label{cmc3}\\
|N(939)\rangle&=&0.90|N^{2}_{8}S_{S}\rangle+0.34|N^{2}_{8}S'_{S}\rangle-0.27|N_{8}^{2}S_{M}\rangle.
\label{cmc4}
\end{eqnarray}
Note that the signs of the second and third coefficients in the above equations are
different from those in Ref.~\cite{Koniuk:1979vy}, due to our definitions for the spin states
$|\frac{1}{2},\pm\frac{1}{2}\rangle_{(\rho,\lambda)}$, the orbital
state $\Phi^{s}_{200}$ and the configuration
$|\Lambda^{2}_{1}P_{A}\rangle$.
We give the explicit wave functions for the components in
$\Lambda(1405)$ in the following two subsections.

\subsection{Wave functions for the three-quark components}
\label{wf3q}
Here we take the flavor-spin-orbital wave functions for the
three-quark components in the considered configurations of
$\Lambda(1405)$ ($\equiv \Lambda^*$) to be of the following forms:
\begin{eqnarray}
|\Lambda(1405)^{2}_{1}P_{A},\frac{1}{2}^{-}\rangle&=&\frac{1}{\sqrt{6}}|\Lambda\rangle_{a}X_{a}\Phi_{\Lambda^{*}}
(\vec{q}_{\lambda},\vec{q}_{\rho})\,,\\
|\Lambda(1405)^{2}_{8}P_{M},\frac{1}{2}^{-}\rangle&=&-\frac{1}{2\sqrt{3}}(|\Lambda\rangle_{\lambda}X_{\lambda}
+|\Lambda\rangle_{\rho}X_{\rho})\Phi_{\Lambda^{*}}(\vec{q}_{\lambda},\vec{q}_{\rho})\,,\\
|\Lambda(1405)^{4}_{8}P_{M},\frac{1}{2}^{-}\rangle&=&\frac{1}{2\sqrt{3}}(|\Lambda\rangle_{\lambda}X_{\lambda}^{'}
+|\Lambda\rangle_{\rho}X_{\rho}^{'})\Phi_{\Lambda^{*}}(\vec{q}_{\lambda},\vec{q}_{\rho})\,,
\end{eqnarray}
where $|\Lambda\rangle_{a}$ and $|\Lambda\rangle_{\rho(\lambda)}$
are the totally anti-symmetric (the flavor singlet) and mixed
symmetric (the flavor octet) flavor wave functions;
$X_a$, $X_{\rho(\lambda)}$, and $X^{\prime}_{\rho(\lambda)}$
denote the completely anti-symmetric and mixed symmetric spin-orbital coupled wave functions, respectively;
$\Phi_{\Lambda^{*}}(\vec{q}_{\lambda},\vec{q}_{\rho})$
the symmetric orbital wave function, and the Jacobi momenta are related to those of the quarks by
 \begin{eqnarray}
 \vec{q}_{\rho}=\frac{1}{\sqrt{2}}(\vec{q}_{1}-\vec{q}_{2}) ~,~ \vec{q}_{\lambda}
 =\frac{1}{\sqrt{6}}(\vec{q}_{1}+\vec{q}_{2}-2\vec{q_{3}}).
 \end{eqnarray}

For the considered configurations of the octet baryons, we employ the following flavor-spin-orbital wave functions:
\begin{eqnarray}
|B^{2}_{8}S_{S},\frac{1}{2}^{+}\rangle_{s_{z}}&=&\frac{1}{\sqrt{2}}(|B\rangle_{\lambda}|\frac{1}{2},s_{z}\rangle_{\lambda}
+|B\rangle_{\rho}|\frac{1}{2},s_{z}\rangle_{\rho})\Phi_{000}(\vec{q}_{\lambda},\vec{q}_{\rho})\,,\\
|B^{2}_{8}S_{S}^{'},\frac{1}{2}^{+}\rangle_{s_{z}}&=&\frac{1}{\sqrt{2}}(|B\rangle_{\lambda}|\frac{1}{2},s_{z}\rangle_{\lambda}
+|B\rangle_{\rho}|\frac{1}{2},s_{z}\rangle_{\rho})\Phi^{s}_{200}(\vec{q}_{\lambda},\vec{q}_{\rho})\,,\\
|B^{2}_{8}S_{M},\frac{1}{2}^{+}\rangle_{s_{z}}&=&\frac{1}{2}[(|B\rangle_{\lambda}|\frac{1}{2},s_{z}\rangle_{\rho}
+|B\rangle_{\rho}|\frac{1}{2},s_{z}\rangle_{\lambda})\Phi^{\rho}_{200}(\vec{q}_{\lambda},\vec{q}_{\rho})-(|B\rangle_{\lambda}|\frac{1}{2},s_{z}\rangle_{\lambda}
+|B\rangle_{\rho}|\frac{1}{2},s_{z}\rangle_{\rho})\nonumber\\
&&\Phi_{200}^{\lambda}(\vec{q}_{\lambda},\vec{q}_{\rho})]\,.
\end{eqnarray}
Here $|B\rangle_{\rho(\lambda)}$ denotes the mixed symmetric flavor wave function for the corresponding baryon,
and $|\frac{1}{2},s_{z}\rangle_{\rho(\lambda)}$ the mixed symmetric spin wave function.
$\Phi_{000}(\vec{q}_{\lambda},\vec{q}_{\rho})$,
$\Phi^{s}_{200}(\vec{q}_{\lambda},\vec{q}_{\rho})$,
$\Phi^{\rho}_{200}(\vec{q}_{\lambda},\vec{q}_{\rho})$ and
$\Phi_{200}^{\lambda}(\vec{q}_{\lambda},\vec{q}_{\rho})$
are the harmonic orbital wave functions with the subscripts being the corresponding $nlm$ quantum numbers.
The explicit forms for all of the flavor, spin, and orbital wave functions are given in Appendix \ref{tq}.

\subsection{Wave functions for the five-quark components}
\label{wf5q}
Flavor-spin-orbital configurations of the four-quark subsystems in the five-quark components,
with lowest energy for the $J^{p}=\frac{1}{2}^{-}$ resonances, are~\cite{An:2008tz,Helminen:2000jb}
$[31]_{FSX}[4]_{X}[31]_{FS}[211]_{F}[22]_{S}$, with the hyperfine interaction between the
quarks (anti-quark) assumed to depend either on flavor and spin \cite{Glozman:1995fu} or on color and spin \cite{Capstick:2000qj}.
Accordingly, the octet baryon is $[31]_{FSX}[31]_{X}[4]_{FS}[22]_{F}[22]_{S}$.

Wave functions for the five-quark components in the $\Lambda(1405)$ resonance, and for the octet baryons
can be written, respectively, in the following general forms:
\begin{eqnarray}
|\Lambda(1405),s_{z}\rangle_{5q}&=&\sum_{abc}C^{[1^4]}_{[31]_{a}
[211]_{a}}C^{[31]_{a}}_{[211]_{b}[22]_{c}}[4]_{X}[211]_{F}(b)[22]_{S}(c)[211]_{C}(a)
\bar \chi_{s_{z}}\Psi(\vec \kappa_i)\, \label{1405wfc},
\end{eqnarray}
\begin{eqnarray}
|B_{octet},s_{z}\rangle_{5q}&=&\sum_{a,b,c}\sum_{m,s}C^{\frac{1}{2}s_{z}}_{1m,
\frac{1}{2}s}C^{[1^{4}]}_{[31]_{a}[211]_{a}} C^{[4]}_{[22]_{b}[22]_{c}}
[211]_{C}(a)[31]_{X,m}(a)[22]_{F}(b)[22]_{S}(c)\bar{\chi}_{s}\nonumber\\
&&\times\psi(\vec \kappa_{i})\,.\label{octetwfc}
\end{eqnarray}
Here the color, space, and flavor-spin wave functions of 4-quark
subsystem are denoted in their Young patterns. The sum over
$a$ runs over the 3 configurations of the $[211]_{C}$ and
$[31]_{XFS}$, those over $b$ and $c$ run over all the configurations of the
$[22]$ and $[211]$ representations of $S_4$, respectively. $C^{[1^4]}_{[31]_{a}
[211]_{a}}$ and $C^{[31]_{a}}_{[211]_{b}[22]_{c}}$ are the
Clebsch-Gordan coefficients of the $S_{4}$ permutation group, the
values of which are $C^{[1^4]}_{[31]_{1} [211]_{1}}$ =
$-~C^{[1^4]}_{[31]_{2} [211]_{2}}$ = $C^{[1^4]}_{[31]_{3}
[211]_{3}}$ = $\frac{1}{\sqrt {3}}$,
$C^{[4]}_{[22]_{b}[22]_{c}}=\frac{1}{\sqrt{2}}\delta_{bc}$ and
the coefficients
$C^{[31]_{a}}_{[211]_{b}[22]_{c}}$ are shown in the decompositions
of the $|[31]_{FS}\rangle$ configurations in Appendix \ref{fqfs}.
The orbital, flavor, spin, and color wave functions are denoted by
the Weyl tableau, and we give the explicit forms for those wave
functions in Appendix B. $\Psi(\vec \kappa_i)$ and
$\psi(\vec\kappa_{i})$ in Eqs. (\ref{1405wfc}) and
(\ref{octetwfc}) are the orbital symmetric wave functions for the
five-quark components in $\Lambda(1405)$ and the octet baryons,
respectively, with the Jacobi momenta
\begin{eqnarray}
\vec{\kappa}_{1}&=&\sqrt{\frac{1}{2}}(\vec{q}_{1}-\vec{q}_{2}),\vec{\kappa}_{2}
=\sqrt{\frac{1}{6}}(\vec{q}_{1}+\vec{q}_{2}-2\vec{q}_{3})\,,\\
\vec{\kappa}_{3}&=&\sqrt{\frac{1}{12}}(\vec{q}_{1}+\vec{q}_{2}+\vec{q}_{3}-3\vec{q}_{4}),\vec{\kappa}_{4}
=\sqrt{\frac{1}{20}}(\vec{q}_{1}+\vec{q}_{2}+\vec{q}_{3}+\vec{q}_{4}-4\vec{q}_{5})\,.
\end{eqnarray}
The orbital configuration $[4]_{X}$ for $\Lambda(1405)$ is
completely symmetric, which means all of the quarks and anti-quark
should be in their orbital ground states, and the explicit form of
the mixed symmetric orbital configuration $[31]_{X}$ for the octet
baryons is
\begin{eqnarray}
|[31]\rangle_{X1}&=&\sqrt{\frac{1}{12}}\{3|0001\rangle-|0010\rangle-|0100\rangle-|1000\rangle\}\,,
\label{31xa}\\
|[31]\rangle_{X2}&=&\sqrt{\frac{1}{6}}\{2|0010\rangle-|0100\rangle-|1000\rangle\}\,,
\label{31xb}\\
|[31]\rangle_{X3}&=&\sqrt{\frac{1}{2}}\{|0100\rangle-|1000\rangle\}\,,
\label{31x}
\end{eqnarray}
where $0$ and $1$ correspond to the quark in its ground or first orbitally
excited state, respectively. The explicit orbital wave
function is the combination of the orbital configuration, Eqs. (\ref{31xa})-(\ref{31x}),
and the symmetric wave function $\psi(\vec\kappa_{i})$.
Explicit color-orbital coupled wave function are reported in
Appendix \ref{fq}.

In Table \ref{tab1} we give for decomposition of baryon states the relevant flavor-spin
configurations, as well as the
Coefficients $A_{i}$, Eq.~\ref{wfb}. The corresponding five-quark components
in $\Lambda(1405)^{2}_{8}P_{M}$ and $B^{2}_{8}S_{S}$ are taken from Ref. \cite{an:2006zf},
and those for the other configurations are obtained by employing the weight diagram
method \cite{ma}.
In this latter case, one can also apply the $SU(3)$ uppering and lowering operators in the
flavor space.

%
\begin{table}[ht!]
\caption{{\footnotesize Five-quark components in
$\Lambda(1405)$, $\Lambda$, $\Sigma^{0}$, the proton and the corresponding coefficients.}}
\begin{tabular}{ccccccc}
\hline \hline
          Baryon   & Flavor-spin configuration &   $A_{u}$ &&   $A_{d}$ && $A_{s}$ \\
\hline

$\Lambda(1405)^{2}_{1}P_{A}$ & $[211]_{F}$ &  $\sqrt{\frac{1}{3}}$  &&
$\sqrt{\frac{1}{3}}$ && $\sqrt{\frac{1}{3}}$   \\

$\Lambda(1405)^{2}_{8}P_{M}$    & $[211]_{F}$    &  $-\sqrt{\frac{1}{6}}$  &&
 $-\sqrt{\frac{1}{6}}$ && $\sqrt{\frac{2}{3}}$ \\

$\Lambda(1405)^{4}_{8}P_{M}$ & $[211]_{F}$ &  $-\sqrt{\frac{1}{6}}$  &&
 $-\sqrt{\frac{1}{6}}$ && $\sqrt{\frac{2}{3}}$ \\

$\Lambda(1116)^{2}_{8}S_{S}$ & $[22]_{F}$ & $-\sqrt{\frac{1}{2}}$  &&
  $-\sqrt{\frac{1}{2}}$ && $0$ \\

$\Lambda(1116)^{2}_{8}S'_{S}$ & $[22]_{F}$&  $-\sqrt{\frac{1}{2}}$ &&
  $-\sqrt{\frac{1}{2}}$ && $0$ \\

$\Lambda(1116)^{2}_{8}S_{M}$  & $[22]_{F}$ &  $-\sqrt{\frac{1}{2}}$ &&
  $-\sqrt{\frac{1}{2}}$ && $0$ \\

$\Sigma(1194)^{2}_{8}S_{S}$ & $[22]_{F}$ &  $-\sqrt{\frac{1}{6}}$ &&
 $\sqrt{\frac{1}{6}}$  && $\sqrt{\frac{2}{3}}$  \\

$\Sigma(1194)^{2}_{8}S'_{S}$ & $[22]_{F}$ &  $-\sqrt{\frac{1}{6}}$ &&
  $\sqrt{\frac{1}{6}}$ && $\sqrt{\frac{2}{3}}$ \\

$\Sigma(1194)^{2}_{8}S_{M}$ & $[22]_{F}$ &  $-\sqrt{\frac{1}{6}}$ &&
 $\sqrt{\frac{1}{6}}$ && $\sqrt{\frac{2}{3}}$ \\

$N(939)^{2}_{8}S_{S}$ & $[22]_{F}$ &  $0$ && $\sqrt{\frac{2}{3}}$ && $\sqrt{\frac{1}{3}}$ \\

$N(939)^{2}_{8}S'_{S}$ & $[22]_{F}$ &  $0$ && $\sqrt{\frac{2}{3}}$ && $\sqrt{\frac{1}{3}}$ \\

$N(939)^{2}_{8}S_{M}$ & $[22]_{F}$ &  $0$ && $\sqrt{\frac{2}{3}}$  && $\sqrt{\frac{1}{3}}$ \\

\hline \hline
\end{tabular}
\label{tab1}
\end{table}
%

\section{Formalism for the radiative and strong decays}
\label{fmm}

Taking into account the five-quark components, the decays of a baryon embodies three types of
possible transitions:
$i)$ between the three-quark,
$ii)$ between the five-quark,
$iii)$ between three- and five-quark.
The first two processes are the so-called diagonal,
and the last one nondiagonal transitions.

In the next two subsections, we describe briefly the formalism for radiative and strong decays of the
baryons in a non-relativistic quark model.

\subsection{Formalism for radiative decay}
\label{frd}
It is established that the radiative decay of baryons can be described by the
helicity amplitudes for the electromagnetic transitions.
For $\gamma^{*}Y\rightarrow\Lambda(1405)$, with $Y \equiv \Lambda(1116),~ \Sigma(1193)$,
they are defined as follows:
\begin{equation}
A_{1/2}=\sqrt{\frac{2\pi\alpha}{K}}\frac{1}{e}\langle
\Lambda(1405),S^{*}_{z}|=\frac{1}{2}
|\epsilon^{+}_{\mu}J^{\mu}|B,S_{z}=-\frac{1}{2}\rangle.
\end{equation}
Here $\epsilon^{+}_{\mu}$ is the polarization vector for the
right-handed photon, $J^{\mu}$ denotes the electromagnetic current,
and $K$ the real photon three-momentum magnitude in the centre-of-mass frame of the
$\Lambda(1405)$ resonance.
For the
$\Lambda(1405) \to \Lambda(1116)\gamma$, $\Sigma(1194)\gamma$ radiative decays, the
values for $K$ are about $259$ MeV/c and $195$ MeV/c, respectively.

The diagonal electromagnetic transition operator in the non-relativistic constituent quark model
takes \cite{Yu:2006sc,Aznauryan:2008us} the following form:
\begin{equation}
\hat{T}_{d}=\sum_{i}^{nq}\sqrt{2}\hat{\mu}_{i}\phi^{i'\dag}_{z}\pmatrix{
\sqrt{2}q_{i+} & k \cr 0 & \sqrt{2}q_{i+}}\phi^{i}_{z}.
\label{radiatived}
\end{equation}
Here the sum over $i$ runs over the quark contents of the corresponding components,
i.e. $nq=3$ for the three-quark and $nq=5$ for the five-quark components.
$\hat{\mu}_{i}=\frac{e_{i}}{2m_{i}}$ denotes the magnetic moment
operator of the $i^{th}$ quark, $\phi^{i'}_{z}$ and $\phi^{i}_{z}$ are
the $i^{th}$ quark spin operators for the initial and final states, respectively, and
$q_{i+}=\frac{1}{\sqrt{2}}(q_{ix}+iq_{iy})$ with $\vec{q}_{i}$ being
the momentum of the $i^{th}$ quark.
Finally, $k$ is the z-component of
the photon momentum. Note that we have taken the photon momentum to
be $\vec{k}=(0,0,k)$, and it is related to the square of the
four-momentum transfer $Q^{2}$
\begin{equation}
k^{2}=Q^{2}+\frac{(M_{\Lambda(1405)}^{2}-m_{Y}^{2}-Q^{2})^{2}}{4M^{2}},
\end{equation}
where ${Y}\equiv \Lambda(1160),~\Sigma(1193)$.

For the nondiagonal transitions, taking the $q\bar{q}-\gamma$
vertices to have the elementary forms

$\bar{u}(q_{i})\gamma^{\mu}v(\bar{q})$ ($3q\rightarrow5q$) and
$\bar{v}(\bar{q})\gamma^{\mu}u(q_{i})$ ($5q\rightarrow3q$), then
the transition operators in the non-relativistic constituent quark model
can be derived
\begin{eqnarray}
\hat{T}_{35}&=&\sum_{i}^{4}\sqrt{2}e_{i}\phi^{i\dag}_{z}\pmatrix{ 0
&
1\cr 0 & 0}\phi^{\bar{q}}_{z}\,,\\ \nonumber \\
\hat{T}_{53}&=&\sum_{i}^{4}\sqrt{2}e_{i}\phi^{\bar{q}\dag}_{z}\pmatrix{
0 &
1\cr 0 & 0}\phi^{i}_{z}\,.
\label{radiativend}
\end{eqnarray}
Here $\hat{T}_{35}$ and $\hat{T}_{53}$ are the operators for the
$\gamma^{*}qqq\rightarrow qqqq\bar{q}$ and
$\gamma^{*}qqqq\bar{q}\rightarrow qqq$ transitions, respectively.

Thus, the helicity amplitude $A_{1/2}$ for the
electromagnetic transition $\gamma^{*}Y\rightarrow\Lambda(1405)$ can be written in the following form:
\begin{equation}
A_{1/2}=\sqrt{\frac{2\pi\alpha}{K}}\frac{1}{e}\langle\Lambda(1405),\frac{1}{2}|(\hat{T}_{d}+\hat{T}_{a})|Y,-\frac{1}{2}\rangle,
\end{equation}
where we have defined $\hat{T}_{a}=\hat{T}_{35}+\hat{T}_{53}$, which correspond to nondiagonal transitions.

\begin{table}[ht]
\caption{\footnotesize Helicity amplitude $A^{\Lambda}_{1/2}$ for
electromagnetic transition $\gamma^{*}\Lambda \to \Lambda(1405)$.
Note that the full amplitudes in columns 2 to 4 are obtained by
multiplying each term by the following expressions:
$\sqrt{\frac{2\pi\alpha}{K}}A_{3q}^{\Lambda}A_{3q}^{\Lambda^{*}}\exp\{-\frac{k^{2}}{6\omega_{3}^{2}}\}$
for $3q\to3q$,
$\sqrt{\frac{2\pi\alpha}{K}}A_{5q}^{\Lambda}A_{5q}^{\Lambda^{*}}\frac{1}{24}
(\frac{1}{m}+\frac{2}{m_{s}})\omega_{5}\exp\{-\frac{k^{2}}{5\omega_{3}^{2}}\}$
for $5q\to5q$, and
$\sqrt{\frac{2\pi\alpha}{K}}A_{3q}^{\Lambda}A_{5q}^{\Lambda^{*}}C_{35}\exp\{-\frac{3k^{2}}{20\omega_{3}^{2}}\}$
for $N-D$. }
\begin{tabular} {lccc}
\hline \hline
& $3q \to 3q$ & $5q \to 5q$ & $N-D$ \\
\hline
$\Lambda^{2}_{8}S_{S}$ $\to$ $\Lambda^{2}_{1}P_{A}$ &
$\frac{1}{18}
(\frac{1}{m}+\frac{2}{m_{s}})\omega_{3}(1+\frac{k^{2}}{2\omega_{3}^{2}})$
& $1/\sqrt{3}$ & $\frac{1}{6}
$ \\
$\Lambda^{2}_{8}S_{S}$ $\to$ $\Lambda^{2}_{8}P_{M}$ &
$\frac{1}{36}[(\frac{1}{m}-\frac{2}{m_{s}})
\frac{k^{2}}{\omega_{3}}-(\frac{1}{m}+\frac{2}{m_{s}})2\omega_{3}]$
& $-1/\sqrt{6}$& $\frac{\sqrt{2}}{12}$  \\
$\Lambda^{2}_{8}S_{S}$ $\to$ $\Lambda^{4}_{8}P_{M}$ & $\frac{1}{36m}
\frac{k^{2}}{\omega_{3}}$ & $-1/\sqrt{6}$ & $\frac{\sqrt{2}}{12}$  \\
$\Lambda^{2}_{8}S'_{S}$ $\to$ $\Lambda^{2}_{1}P_{A}$ &
$-\frac{1}{18\sqrt{3}}
(\frac{1}{m}+\frac{2}{m_{s}})\omega_{3}[(1+\frac{k^{2}}{6\omega_{3}^{2}})
-(1-\frac{k^{2}}{6\omega_{3}^{2}})\frac{k^{2}}{2\omega_{3}^{2}}]$ &
$1/\sqrt{3}$ & 0  \\
$\Lambda^{2}_{8}S'_{S}$ $\to$ $\Lambda^{2}_{8}P_{M}$ &
$\frac{1}{54\sqrt{3}}[(\frac{1}{2m}-\frac{1}{m_{s}})
\frac{k^{2}}{\omega_{3}}(1-\frac{k^{2}}{6\omega_{3}^{2}})+(\frac{1}{m}-\frac{2}{m_{s}})2\omega_{3}
(1+\frac{k^{2}}{6\omega_3^{2}})]$ & $-1/\sqrt{6}$ & 0  \\
$\Lambda^{2}_{8}S'_{S}$ $\to$ $\Lambda^{4}_{8}P_{M}$ &
$\frac{1}{36\sqrt{3}m}
(1-\frac{k^{2}}{6\omega^{2}_{3}})\frac{k^{2}}{\omega_{3}}$ & $-1/\sqrt{6}$ & 0  \\
$\Lambda^{2}_{8}S_{M}$ $\to$ $\Lambda^{2}_{1}P_{A}$ &
-$\frac{\sqrt{6}}{54}
(\frac{1}{m}+\frac{2}{m_{s}})\omega_{3}(1-\frac{k^{2}}{12\omega_{3}^{2}}+\frac{k^{4}}{24\omega_{3}^{4}})$
& $1/\sqrt{3}$ & 0  \\
$\Lambda^{2}_{8}S_{M}$ $\to$ $\Lambda^{2}_{8}P_{M}$ &
$-\frac{\sqrt{6}}{108}\omega_{3}[(\frac{1}{m}+\frac{1}{m_{s}})
\frac{k^{2}}{\omega_{3}^{2}}-\frac{k^{4}}{6m_{s}\omega_{3}^{4}}]$ &
$-1/\sqrt{6}$
 & 0  \\
$\Lambda^{2}_{8}S_{M}$ $\to$ $\Lambda^{4}_{8}P_{M}$ &
$-\frac{\sqrt{6}}{162}
\omega_{3}[(\frac{1}{m}-\frac{1}{m_{s}})\frac{k^{2}}{\omega_{3}^{2}}-\frac{k^{4}}{8m\omega_{3}^{4}}]$
& $-1/\sqrt{6}$ & 0  \\
\hline
\end{tabular}
\label{alam}
\end{table}
%

Taking into account the configurations mixing effects and the contributions of the five-quark
components, we need to calculate $36$ transition amplitudes for each decay.
For the diagonal transitions ($3q \to 3q$ and $5q \to 5q$)  the
calculations are similar to that in Refs.
\cite{Koniuk:1979vy,Capstick:2000qj,close}. Explicit calculations of the
nondiagonal ($N-D$) electromagnetic transitions elements in our approach are similar
to the one in Ref. \cite{An:2008xk} for the
$\gamma^{*}N\rightarrow N(1535)$ process.
Amplitudes for $\gamma^{*} \Lambda(1116) \to \Lambda(1405)$ and
$\gamma^{*} \Sigma^\circ(1194) \to \Lambda(1405)$
are given in Tables \ref{alam} and \ref{asig}, respectively.
%
\begin{table}[ht]
\caption{\footnotesize Helicity amplitude $A^{\Sigma}_{1/2}$ for
electromagnetic transition $\gamma^{*}\Sigma^{0}\to \Lambda(1405)$.
Note that the full amplitudes in columns 2 to 4 are obtained by
multiplying each term by the following expressions:
$\sqrt{\frac{2\pi\alpha}{K}}A_{3q}^{\Sigma}A_{3q}^{\Lambda^{*}}\exp\{-\frac{k^{2}}{6\omega_{3}^{2}}\}$
for $3q \to 3q$,
$\sqrt{\frac{2\pi\alpha}{K}}A_{5q}^{\Sigma}A_{5q}^{\Lambda^{*}}\frac{2\omega_{5}}{m}\exp\{-\frac{k^{2}}{5\omega_{3}^{2}}\}$
for $5q \to 5q$, and
$\sqrt{\frac{2\pi\alpha}{K}}A_{3q}^{\Sigma}A_{5q}^{\Lambda^{*}}C_{35}\exp\{-\frac{3k^{2}}{20\omega_{3}^{2}}\}$
for $N-D$.}
\begin{tabular} {lccc}
\hline \hline
& $3q \to 3q$ & $5q \to 5q$ & $N-D$ \\
\hline
$\Sigma^{2}_{8}S_{S}~\to~\Lambda^{2}_{1}P_{A}$ &
$-\frac{1}{2\sqrt{3}}
\frac{\omega_{3}}{m}(1+\frac{k^{2}}{2\omega_{3}^{2}})$ &
$-\frac{1}{16}$ & $-\frac{1}{2\sqrt{3}}$ \\
$\Sigma^{2}_{8}S_{S}~\to~\Lambda^{2}_{8}P_{M}$ &
$-\frac{1}{4\sqrt{3}}
\frac{\omega_{3}}{m}(2+\frac{k^{2}}{3\omega_{3}^{2}})$
& $-\frac{3}{16\sqrt{2}}$& $\frac{1}{2 \sqrt{6}}$  \\
$\Sigma^{2}_{8}S_{S}~\to~\Lambda^{4}_{8}P_{M}$ &
$\frac{\sqrt{3}}{36m}\frac{k^{2}}{\omega_{3}}$
& $-\frac{3}{16\sqrt{2}}$ & $\frac{1}{2 \sqrt{6}}$  \\
$\Sigma^{2}_{8}S'_{S}~\to~\Lambda^{2}_{1}P_{A}$ &
$\frac{\omega_{3}}{6m}
[(1+\frac{k^{2}}{6\omega_{3}^{2}})-(1-\frac{k^{2}}{6\omega_{3}^{2}})\frac{k^{2}}{2\omega_{3}^{2}}]$
&
$-\frac{1}{16}$ & 0  \\
$\Sigma^{2}_{8}S'_{S}~\to~\Lambda^{2}_{8}P_{M}$ & $-\frac{1}{4m}[
\frac{k^{2}}{9\omega_{3}}(1-\frac{k^{2}}{6\omega_{3}^{2}})-\frac{2\omega_{3}}{3}(1+\frac{k^{2}}{6\omega_{3}^{2}})]$
&
$-\frac{3}{16\sqrt{2}}$ & 0  \\
$\Sigma^{2}_{8}S'_{S}~\to~\Lambda^{4}_{8}P_{M}$ & $\frac{1}{36m}
(1-\frac{k^{2}}{6\omega^{2}_{3}})\frac{k^{2}}{\omega_{3}}$ & $-\frac{3}{16\sqrt{2}}$ & 0  \\
$\Sigma^{2}_{8}S_{M}~\to~\Lambda^{2}_{1}P_{A}$ &
$\frac{\sqrt{2}}{6}
\frac{\omega_{3}}{m}(1-\frac{k^{2}}{12\omega_{3}^{2}}+\frac{k^{4}}{24\omega_{3}^{4}})$
& $-\frac{1}{16}$ & 0  \\
$\Sigma^{2}_{8}S_{M}~\to~\Lambda^{2}_{8}P_{M}$ &
$-\frac{\sqrt{2}}{72}\frac{\omega_{3}}{m}[
\frac{k^{2}}{\omega_{3}^{2}}-\frac{k^{4}}{6\omega_{3}^{4}}]$ &
$-\frac{3}{16\sqrt{2}}$
 & 0  \\
$\Sigma^{2}_{8}S_{M}~\to~\Lambda^{4}_{8}P_{M}$ &
$-\frac{\sqrt{2}\omega_{3}}{72m} \frac{k^{4}}{\omega_{3}^{4}}$
& $-\frac{3}{16\sqrt{2}}$ & 0  \\
\hline
\end{tabular}
\label{asig}
\end{table}
%

Notice that, in Tables \ref{alam} and \ref{asig} we have defined
\begin{equation}
C_{35}=\langle\varphi_{00}(\vec{\kappa}_{1})\varphi_{00}(\vec{\kappa}_{2})
|\varphi_{00}(\vec{\kappa}_{1})\varphi_{00}(\vec{\kappa}_{2})\rangle\
=(\frac{2\omega_{3}\omega_{5}}{\omega_{3}^{2}+\omega_{5}^{2}})^{3}
\, ,
\end{equation}
which is the orbital overlap integral factor in the matrix
elements of the nondiagonal transitions. Here, $\omega_{3}$ and
$\omega_{5}$ are the oscillator frequencies for the $qqq$ and
$qqqq\bar{q}$ systems, respectively.

Finally,
the radiative decay width of $\Lambda(1405)$ in terms of the helicity
amplitudes $A_{1/2}$ at the real photon point is \cite{close}
\begin{equation}
\Gamma_{Y\gamma}=\frac{k^{2}m_{Y}}{\pi M}|A_{1/2}(Q^{2}=0)|^{2}.
\label{gammar}
\end{equation}

\subsection{Formalism for strong decay}
\label{fsd}

In the chiral constituent quark model, the coupling of the light
quarks ($u,d,s$) to the octet of light pseudoscalar mesons takes the
form
\begin{equation}
\mathcal{L}_{Mqq}=i\frac{g^{q}_{A}}{2f_{M}}\bar{\psi}_{q}\gamma_{5}\gamma_{\mu}
\partial^{\mu}m_{a}\lambda_{a}\psi_{q}\,.\label{lag1}
\end{equation}
Here, $g^{q}_{A}$ denotes the axial coupling constant for the constituent quarks,
$f_{M}$ is the decay constant of meson $M$ ($\pi$, $K$).
$\psi_{q}$ is the quark
field and $m_{a}$ the meson field. Finally, $\lambda_{a}$s are the
$SU(3)$ Gell-Mann matrices. Combination of Eq. (\ref{lag1})
with the representation
\begin{eqnarray}
m_{a}\lambda_{a}&=&\sqrt{2}\pmatrix{ \frac{1}{\sqrt{2}}\pi^{\circ}+\frac{1}{\sqrt{6}}\eta
& \pi^{+} & K^{+} \cr
\pi^{-}   & -\frac{1}{\sqrt{2}}\pi^{\circ}+\frac{1}{\sqrt{6}}\eta  & K^{\circ} \cr
{K}^{-}& \bar{K}^{\circ} & -\sqrt{\frac{2}{3}}\eta}.
\end{eqnarray}
leads to the following quark-meson-quark chiral coupling in the
momentum space
\begin{equation}
\mathcal{L}_{Mqq}=\frac{g^{q}_{A}}{2f_{M}}\bar{\psi}_{q}\gamma_{5}\gamma_{\mu}
k^{\mu}X_{M}^{q}\psi_{q}\,,\label{lag2}
\end{equation}
where $X_{M}^{q}$ is the flavor operator for emission of meson $M$
from the corresponding quark $q$
\begin{eqnarray}
X_{\pi^{\circ}}^{q}&=&\lambda_{3}\,,\\
X_{K^{\pm}}^{q}&=&\frac{1}{\sqrt{2}}(\lambda_{4}+\lambda_{5})\,,\\
X_{K^{\circ}}^{q}&=&-{\frac{1}{\sqrt 2}}(\lambda_{6}-i\lambda_{7})\,,\\ X_{\eta}^{q}&=&cos\theta\lambda_{8}-sin\theta\mathcal{I},
X_{\eta'}^{q}=cos\theta\lambda_{8}+sin\theta\mathcal{I},
\end{eqnarray}
with $\mathcal{I}$ the unit operator in the $SU(3)$ flavor space.

Within the non-relativistic approximation, we can get the
baryon-meson-baryon coupling in the chiral constituent quark model
\begin{eqnarray}
\hat{T}^{M}_{d}&=&\sum_{i}^{nq}\frac{g_{A}^{q}}{2f_{M}}\phi^{i'\dag}_{z}\pmatrix{(1+\frac{k_{0}}{2m_{f}})k_M
-\frac{k_{0}}{2\mu}q_{iz}
& -\sqrt{2}\frac{k_{0}}{2\mu} q_{i-} \cr
-\sqrt{2}\frac{k_{0}}{2\mu} q_{i+} &
-(1+\frac{k_{0}}{2m_{f}})k_{M}+\frac{k_{0}}{2\mu}q_{iz}}\phi^{i}_{z}X_{M}^{i}\,,\\ \nonumber \\
%
\hat{T}^{M}_{53}&=&-\sum_{i}^{4}\frac{g_{A}^{q}}{2f_{M}}(m_{i}+m_{f})\phi^{\bar{q}\dag}_{z}
\pmatrix{1&0\cr0&1}\phi^{i}_{z}X^{i}_{M}\,,\\ \nonumber \\
%
\hat{T}^{M}_{35}&=&-\sum_{i}^{4}\frac{g_{A}^{q}}{2f_{M}}(m_{i}+m_{f})\phi^{i\dag}_{z}\pmatrix{1&0\cr0&1}\phi^{\bar{q}}_{z}X^{i}_{M}.
\end{eqnarray}
Here $m_{i}$ and $m_{f}$ are the initial and final constituent
masses of the quark which emits a meson, and
$\mu={m_{i}m_{f}}/({m_{i}+m_{f}})$. $k_{0}$ and $k_{M}$ denote
the energy and magnitude of the three-momentum of the final meson in
the centre-of-mass frame of the initial baryon.
Note that we have taken the meson three-momentum  to be in the z-direction,
$\vec{k_{M}}=(0,0,k_{M})$, and it is related to the masses of the initial
and final hadrons
\begin{eqnarray}
k_{M}&=&\{[M^{2}_{i}-(M_{f}+m_{M})^{2}][M^{2}_{i}-(M_{f}
-m_{M})^{2}]\}^{1/2}/2M_{i}\,.\
\end{eqnarray}
The transition amplitudes are obtained by the calculations of the
following matrix elements
\begin{equation}
T^{M}=\langle\Lambda(1405),\frac{1}{2}|(\hat{T}^{M}_{d}+\hat{T}^{M}_{a})|Y,\frac{1}{2}\rangle,
\label{TM}
\end{equation}
where we have defined $\hat{T}^{M}_{a}=\hat{T}^{M}_{35}+\hat{T}^{M}_{53}$.

\begin{table}[ht]
\caption{\footnotesize Transition amplitudes of the $\Lambda(1405) \to \Sigma(1194) \pi$ decay. Note that
the full amplitudes in columns 2 and 3 are obtained by multiplying each term by the
following expressions:
$\frac{g}{2f_{\pi}}A^{\Sigma^{\circ}}_{3q}A^{\Lambda^{*}}_{3q}\omega_{3}\exp\{-\frac{k^{2}}{6\omega_{3}^{2}}\}$
for column $3q \to 3q$, and
$\frac{g}{f_{\pi}}A^{\Sigma^{\circ}}_{3q}A^{\Lambda^{*}}_{5q}
mC_{35}\exp\{-\frac{3k^{2}}{20\omega_{5}^{2}}\}$ for column
$N-D$. Here $k$ denotes the $\pi$ three-momentum magnitude
$k_{\pi}$, and $k_{0}$ the energy of the $\pi$ meson.}
\begin{tabular} {lcc}
\hline \hline
& $3q \to 3q$ & N-D \\
\hline
$\Lambda^{2}_{1}P_{A}~\to~\Sigma^{2}_{8}S_{S}$    &
$-\frac{1}{\sqrt{6}}
[(1+\frac{k_{0}}{6m})\frac{k^{2}}{\omega_{3}^{2}}-3\frac{k_{0}}{m}]$
& $\frac{1}{\sqrt{6}}$    \\
$\Lambda^{2}_{8}P_{M}~\to~\Sigma^{2}_{8}S_{S}$    &
$-\frac{1}{3\sqrt{6}}
[(1+\frac{k_{0}}{6m})\frac{k^{2}}{\omega_{3}^{2}}-3\frac{k_{0}}{m}]$    & $-\frac{1}{2\sqrt{3}}$    \\
$\Lambda^{4}_{8}P_{M}~\to~\Sigma^{2}_{8}S_{S}$    &
$-\frac{2}{3 \sqrt{6}}
[(1+\frac{k_{0}}{6m})\frac{k^{2}}{\omega_{3}^{2}}-3\frac{k_{0}}{m}]$    & $-\frac{1}{2\sqrt{3}}$   \\
$\Lambda^{2}_{1}P_{A}~\to~\Sigma^{2}_{8}S'_{S}$   &
$-\frac{1}{3\sqrt{2}}
[3\frac{k_{0}}{m}+(1+\frac{k_{0}}{m})\frac{k^{2}}{\omega_{3}^{2}}-(1+\frac{k_{0}}{6m})\frac{k^{4}}{6\omega_{3}^{4}}]$   & 0   \\
$\Lambda^{2}_{8}P_{M}~\to~\Sigma^{2}_{8}S'_{S}$   &
$-\frac{1}{9\sqrt{2}}
[3\frac{k_{0}}{m}+(1+\frac{k_{0}}{m})\frac{k^{2}}{\omega_{3}^{2}}-(1+\frac{k_{0}}{6m})\frac{k^{4}}{6\omega_{3}^{4}}]$   & 0   \\
$\Lambda^{4}_{8}P_{M}~\to~\Sigma^{2}_{8}S'_{S}$   &
$-\frac{\sqrt{2}}{9}
[3\frac{k_{0}}{m}+(1+\frac{k_{0}}{m})\frac{k^{2}}{\omega_{3}^{2}}-(1+\frac{k_{0}}{6m})\frac{k^{4}}{6\omega_{3}^{4}}]$   & 0   \\
$\Lambda^{2}_{1}P_{A}~\to~\Sigma^{2}_{8}S_{M}$    & $-\frac{1}{9}
[3\frac{k_{0}}{m}+(1+\frac{5k_{0}}{12m})\frac{k^{2}}{\omega_{3}^{2}}-(1+\frac{k_{0}}{6m})\frac{k^{4}}{4\omega_{3}^{4}}]$   & 0   \\
$\Lambda^{2}_{8}P_{M}~\to~\Sigma^{2}_{8}S_{M}$    & $-\frac{1}{18}
[3\frac{k_{0}}{m}+(1+\frac{k_{0}}{2m})\frac{k^{2}}{\omega_{3}^{2}}-(1+\frac{k_{0}}{6m})\frac{k^{4}}{3\omega_{3}^{4}}]$   & 0   \\
$\Lambda^{4}_{8}P_{M}~\to~\Sigma^{2}_{8}S_{M}$    & $\frac{1}{9}
[(5+\frac{11k_{0}}{6m})\frac{k^{2}}{6\omega_{3}^{2}}-(1+\frac{k_{0}}{6m})\frac{k^{4}}{6\omega_{3}^{4}}]$   & 0   \\
\hline \hline
\end{tabular}
\label{sigpi}
\end{table}
%

Tables \ref{sigpi} and \ref{kn} give the transition amplitudes for strong decay channel.
We note that none of the diagonal transitions of the five-quark components contributes to the
transition amplitudes.
Those null values can easily be understood, noticing that the
spin configurations for $\Lambda(1405)$ and for the octet baryons are taken to be
$[22]_{S}$, for which the total spin is $S=0$, and there are no
spin-independent terms in the diagonal transition operator (which is not the case
in the electromagnetic transition operator).  However, the configuration mixing effects might
be significant.

\begin{table}[ht]
\caption{\footnotesize Transition amplitudes of the $\Lambda(1405) \to K^{-}p$ decay. Note that
the full amplitudes in columns 2 and 3 are obtained by multiplying each term by the
following expressions:
$\frac{g}{2f_{K}}A^{N}_{3q}A^{\Lambda^{*}}_{3q}\omega_{3}
\exp\{-\frac{k^{2}}{6\omega_{3}^{2}}\}$ for column $3q\to 3q$, and the
factors $\frac{g}{f_{K}}A^{N}_{3q}A^{\Lambda^{*}}_{5q}
(m+m_{s})C_{35}\exp\{-\frac{3k^{2}}{20\omega_{5}^{2}}\}$ for column
$N-D$. Here $k$ denotes the three-momentum magnitude
$k_{K}$, and $k_{0}$ the energy of the $K$ meson.}
\begin{tabular} {lcc}
\hline \hline
& $3q \to 3q$ & $N-D$ \\
\hline $\Lambda^{2}_{1}P_{A}~\to~N^{2}_{8}S_{S}$    &
$-\frac{1}{\sqrt{6}}
[(1+\frac{k_{0}}{2m}-\frac{k_{0}}{6\mu})\frac{k^{2}}{\omega_{3}^{2}}-\frac{3k_{0}}{2\mu}]$
& $-\frac{1}{2 \sqrt{6}}$   \\
$\Lambda^{2}_{8}P_{M}~\to~N^{2}_{8}S_{S}$    & $\frac{1}{\sqrt{6}}
[(1+\frac{k_{0}}{2m}-\frac{k_{0}}{6\mu})\frac{k^{2}}{\omega_{3}^{2}}-3\frac{k_{0}}{2\mu}]$    & $-\frac{1}{4\sqrt{3}}$    \\
$\Lambda^{4}_{8}P_{M}~\to~N^{2}_{8}S_{S}$    & 0         & $-\frac{1}{4\sqrt{3}}$   \\
$\Lambda^{2}_{1}P_{A}~\to~N^{2}_{8}S'_{S}$   &
$-\frac{1}{3\sqrt{2}}
[\frac{3k_{0}}{2\mu}+(1+\frac{k_{0}}{2m}+\frac{k_{0}}{4\mu})\frac{k^{2}}{\omega_{3}^{2}}-(1+\frac{k_{0}}{2m}
-\frac{k_{0}}{6\mu})\frac{k^{4}}{6\omega_{3}^{4}}]$     & 0   \\
$\Lambda^{2}_{8}P_{M}~\to~N^{2}_{8}S'_{S}$   & $\frac{1}{3\sqrt{2}}
[\frac{3k_{0}}{2\mu}+(1+\frac{k_{0}}{2m}+\frac{k_{0}}{4\mu})\frac{k^{2}}{\omega_{3}^{2}}-(1+\frac{k_{0}}{2m}
-\frac{k_{0}}{6\mu})\frac{k^{4}}{6\omega_{3}^{4}}]$    & 0    \\
$\Lambda^{4}_{8}P_{M}~\to~N^{2}_{8}S'_{S}$   & 0         & 0         \\
$\Lambda^{2}_{1}P_{A}~\to~N^{2}_{8}S_{M}$    & $-\frac{1}{9}
[\frac{3k_{0}}{2\mu}+(1+\frac{k_{0}}{2m}-\frac{k_{0}}{24\mu})\frac{k^{2}}{\omega_{3}^{2}}
-(1+\frac{k_{0}}{2m}-\frac{k_{0}}{6\mu})\frac{k^{4}}{4\omega_{3}^{4}}]$    & 0         \\
$\Lambda^{2}_{8}P_{M}~\to~N^{2}_{8}S_{M}$    & $\frac{1}{9}
[\frac{3k_{0}}{\mu}+(2+\frac{k_{0}}{m}-\frac{5k_{0}}{24\mu})\frac{k^{2}}{\omega_{3}^{2}}-(1+\frac{k_{0}}{2m}
-\frac{k_{0}}{6\mu})\frac{k^{4}}{4\omega_{3}^{4}}]$    & 0   \\
$\Lambda^{4}_{8}P_{M}~\to~N^{2}_{8}S_{M}$    & $\frac{2}{9}
[\frac{3k_{0}}{2\mu}+(1+\frac{k_{0}}{2m}-\frac{k_{0}}{6\mu})\frac{k^{2}}{\omega_{3}^{2}}]$    & 0    \\
\hline \hline
\end{tabular}
\label{kn}
\end{table}

 Finally, following Eq. (\ref{TM}), the strong decay width for
$\Lambda(1405) \to (\Sigma(1194)\pi)^\circ$ reads
\begin{equation}
\Gamma_{\Lambda(1405)\rightarrow (\Sigma \pi)^\circ}=\frac{3}{4\pi}\frac{E^{'}+m_{\Sigma}}{M}|\vec{k}_{\pi}|
|T^{\pi}|^{2},
\label{width}
\end{equation}
where $E^{'}$ is the energy of the final $\Sigma$ hyperon
\begin{equation}
E^{'}=\frac{M^{2}-m_{\pi}^{2}+m_{\Sigma}^{2}}{2M}.
\end{equation}
In addition, taking the hadronic level Lagrangian for the
$\Lambda(1405) B M$ coupling, with $B \equiv \Sigma,~N$ and $M \equiv \pi,~K$,  to
be of the following form:
\begin{equation}
\mathcal{L}_{\Lambda(1405)BM}=i\frac{f_{\Lambda(1405)BM}}{m_{M}}\bar{\psi}_{B}\gamma_{\mu}
\partial^{\mu}\phi_{M}X_{M}\psi_{\Lambda(1405)}+h.c.,
\end{equation}
the transition coupling amplitude reads
${f_{\Lambda(1405)BM}(M_{\Lambda(1405)}-m_{B})}/{m_{M}}$.
Comparing the latter expression to the
results obtained in the chiral quark model, one gets
\begin{equation}
\frac{f_{\Lambda(1405)BM}}{m_{M}}=\frac{\langle[\hat{T}^{M}_{d}+\hat{T}^{M}_{35}+\hat{T}^{M}_{53}]\rangle}
{M_{\Lambda(1405)}-m_B}.
\end{equation}

\section{Numerical results and discussion}
\label{nr}
Using the formalism developed in the previous section, here we present
our numerical results for both electromagnetic and strong decays.
Those results have been obtained with no adjustable parameters.
In Table \ref{params} we give the input parameters used in our calculations and
comment on the adopted values.

\begin{table}[ht]
\caption{{\footnotesize The input values used in this manuscript for non vanishing
five-quark probability ($P_{5q}\not=0$).
Here, $m \equiv m_{u}=m_{d}$. For $P_{5q}=0$ we used $m=$340 MeV.
Values in columns 1 to 6 are in MeV.}}
\begin{tabular}{ccccccccccc}
\hline \hline
 $m$ & $m_{s}$ & $\omega_{3}$ & $\omega_{5}$ & $f_{\pi}$ & $f_{K}$ & $g^q_A$
  & $A_{3q}^B$ &  $A_{5q}^{B}$
  & $A_{3q}^{\Lambda ^*}$ & $A_{5q}^{\Lambda^*}$  \\
\hline

 290 & 430     &    340       &    600       &  93       &   113  &  0.82
 & $\sqrt{0.80}$ & $\sqrt{0.20}$ & $\sqrt{0.55}$ & $\sqrt{0.45}$  \\

\hline \hline
\end{tabular}
\label{params}
\end{table}

Since we have introduced the five-quark components in the
baryons,
The constituent quark masses are slightly different
from those used in the traditional constituent quark models.
We take $m_{u}=m_{d}=290$ MeV and $m_{s}=430$ MeV, as suggested in Ref. \cite{An:2009uv}
in order to reproduce the mass of the proton with 20\% five-quark components, and
to investigate successfully the electromagnetic transitions
$\gamma^{*}N \to N^{*}(1535)$ and the strong decays of $N(1535)$.
Values for the oscillator parameters $\omega_{3}$ and $\omega_{5}$ come also from this latter Reference.

The probability of five-quark components in proton leading to
$A_{5q}^{N}$=$\sqrt{0.20}$ (see e.g. Ref.  \cite{An:2009uv}) is also
used for the lowest mass hyprons, $A_{5q}^{\Lambda}$ and $A_{5q}^{\Sigma^\circ}$.
Then the probabilities for $3q$ components are obtained within the used truncated Fock space, implying $(A_{3q}^Y)^2+(A_{5q}^Y)^2$=1. For the $\Lambda$(1405), our numerical results reported
below (see sec. \ref{nr}) favor
$A_{5q}^{\Lambda^*}=\sqrt{0.45}$, and hence, $A_{3q}^{\Lambda^*}=\sqrt{0.55}$.

In Table \ref{params}, $g^{q}_{A}$ denotes the axial coupling constant for the constituent quarks,
and its extracted phenomenological values are \cite{Goity:1998jr,Riska:2000gd,Lahde:2002fe}
in the range $0.70-1.26$.
Here, we have taken $g^{q}_{A}=0.82$, which differs slightly from its value (0.88)
in Ref. \cite{Riska:2000gd} , due to the fact that we have introduced the five-quark
components.

Finally, for  the decay constants of mesons, the empirical values are used
($f_{\pi}=93$ MeV and $f_{K}=113$ MeV).
%
\subsection{Radiative decays of $\Lambda(1405)$}
\label{nrrd}
Helicity amplitudes $A^{\Lambda}_{1/2}$ and $A^{\Sigma}_{1/2}$ for
the electromagnetic transitions $\gamma \Lambda(1116) \to
\Lambda(1405)$ and $\gamma \Sigma(1194) \to \Lambda(1405)$ at the
real photon point are given in Tables \ref{al} and \ref{as},
respectively, showing that the configurations mixing effects are
very important, and the diagonal transitions between the five-quark
components also have non negligible contributions to the helicity
amplitudes. Moreover, as we can see in the nondiagonal
($N-D$) columns, those transitions between the three- and five-quark
components in $Y^{2}_{8}S_{S}$ and $\Lambda(1405)^{2}_{1}P_{A}$
contribute significantly to the helicity amplitudes $A^{Y}_{1/2}$:
about $27\%$ to $A^{\Lambda}_{1/2}$ and $24\%$ to
$A^{\Sigma}_{1/2}$.

Table \ref{rdw} shows our results for the radiative decay widths
of $\Lambda(1405)$, employing Eq. (\ref{gammar}).
Column A contains the results obtained without five-quark
admixture, i.e. $P_{5q}=0\%$, columns B, C, D, and E correspond to $P_{5q}=25\%$, $45\%$,
 $75\%$ and $100\%$, respectively.
The $\Lambda(1405) \to \Lambda \gamma$ channel shows a significant sensitivity ($\approx$ 30\%)
to the five-quark components, roughly in the range 20\%$\lesssim P_{5q}\lesssim$50\%.
For the $\Lambda(1405) \to \Sigma \gamma$ decay, in going from $P_{5q}=0\%$ to $P_{5q}=25\%$, the
decay width increases by roughly 36\%, and drops down with the increasing $P_{5q}$ faster than the
width for $\Lambda(1405) \to \Lambda \gamma$ decay.

%
\begin{table}[ht]
\caption{\footnotesize Results for the helicity amplitude $A^{\Lambda}_{1/2}$ (in
GeV$^{-1/2}$) for electromagnetic transition
$\gamma \Lambda\rightarrow\Lambda(1405)$.}
\begin{tabular} {lccccc}
\hline \hline
& $3q \to 3q$ & $5q \to 5q$ & N-D & total \\
\hline
$\Lambda^{2}_{8}S_{S}~\to~\Lambda^{2}_{1}P_{A}$ & 0.050 & 0.013 & 0.024 & 0.087  \\
$\Lambda^{2}_{8}S_{S}~\to~\Lambda^{2}_{8}P_{M}$ & -0.027 & -0.005& 0.011  & -0.021 \\
$\Lambda^{2}_{8}S_{S}~\to~\Lambda^{4}_{8}P_{M}$ & 0.011 & -0.004 & 0.009  & 0.016 \\

$\Lambda^{2}_{8}S'_{S}~\to~\Lambda^{2}_{1}P_{A}$ & -0.023 & 0.015 & 0  & -0.008 \\
$\Lambda^{2}_{8}S'_{S}~\to~\Lambda^{2}_{8}P_{M}$ & -0.005 & -0.020 & 0  & -0.025 \\
$\Lambda^{2}_{8}S'_{S}~\to~\Lambda^{4}_{8}P_{M}$ & 0.002 & -0.011 & 0  & -0.009 \\

$\Lambda^{2}_{8}S_{M}~\to~\Lambda^{2}_{1}P_{A}$ & -0.019 & 0.008 & 0  & -0.011 \\
$\Lambda^{2}_{8}S_{M}~\to~\Lambda^{2}_{8}P_{M}$ & -0.003 & -0.013 & 0  & -0.016 \\
$\Lambda^{2}_{8}S_{M}~\to~\Lambda^{4}_{8}P_{M}$ & ~0 & -0.053 & 0 &  -0.053 \\
\hline \hline
\end{tabular}
\label{al}
\end{table}
%
%
\begin{table}[ht]
\caption{\footnotesize Results for
the helicity amplitude $A^{\Sigma}_{1/2}$ (in GeV$^{-1/2}$)
of electromagnetic transitions
$\gamma \Sigma\rightarrow\Lambda(1405)$.}
\begin{tabular} {lccccc}
\hline \hline
& $3q ~\to~ 3q$ & $5q ~\to~ 5q$ & N-D & total \\
\hline
$\Sigma^{2}_{8}S_{S}~\to~\Lambda^{2}_{1}P_{A}$ & -0.120 & -0.035 & -0.050 & -0.205 \\
$\Sigma^{2}_{8}S_{S}~\to~\Lambda^{2}_{8}P_{M}$ & -0.073 & -0.041 & 0.021 & -0.093 \\
$\Sigma^{2}_{8}S_{S}~\to~\Lambda^{4}_{8}P_{M}$ & 0.017 & -0.031 & 0.017 & 0.003 \\
$\Sigma^{2}_{8}S'_{S}~\to~\Lambda^{2}_{1}P_{A}$ & 0.038 & -0.027 & 0 & 0.011 \\
$\Sigma^{2}_{8}S'_{S}~\to~\Lambda^{2}_{8}P_{M}$ & 0.278 & -0.353 & 0 & -0.075 \\
$\Sigma^{2}_{8}S'_{S}~\to~\Lambda^{4}_{8}P_{M}$ & 0.002 & -0.093 & 0 & -0.091 \\
$\Sigma^{2}_{8}S_{M}~\to~\Lambda^{2}_{1}P_{A}$ & 0.046 & -0.019 & 0 & 0.027 \\
$\Sigma^{2}_{8}S_{M}~\to~\Lambda^{2}_{8}P_{M}$ & -0.001 & -0.107 & 0 & -0.108 \\
$\Sigma^{2}_{8}S_{M}~\to~\Lambda^{4}_{8}P_{M}$ & 0 & -0.204 & 0 & -0.204 \\
\hline \hline
\end{tabular}
\label{as}
\label{table1}
\end{table}
%

%
%
\begin{table}[ht]
\caption{\footnotesize Results for
the radiative decays widths of $\Lambda(1405)\to {\Lambda(1116) \gamma}$
($\Gamma_{\Lambda \gamma}$), $\Lambda(1405)\to \Sigma(1194)\gamma$
($\Gamma_{\Sigma \gamma }$) (in keV), and their ratio
$R=\Gamma_{\Sigma \gamma} / \Gamma_{\Lambda \gamma}$.}
\begin{tabular} {lccccc}
\hline \hline
& A & B & C & D & E \\
$P_{5q}$ (\%)& 0 & 25 & 45 & 75 & 100 \\
 \hline

$\Gamma_{\Lambda \gamma}$ & 91 & 122 &  123 & 104 & 56  \\

$\Gamma_{\Sigma \gamma}$ & 164 & 223 & 212 & 164 & 73  \\

$R$ & 1.8 & 1.8 & 1.7 & 1.6 & 1.3  \\

\hline \hline
\end{tabular}
\label{rdw}
\end{table}

Table \ref{others} summarizes the widths for the electromagnetic decay of the $\Lambda(1405)$
reported by several authors.
One of the early extractions of those quantities is due to Burkhardt and Lowe~\cite{Burkhardt:1991ms},
motivated by the advent of reliable $K^-p$ atom data \cite{Whitehouse:1989yi} published in late 80's.
Since then, those results have been introduced in PDG, and are considered by some authors as "`data"',
though Burkhardt and Lowe state clearly in their paper the highly phenomenological character of their
investigation, e.g.
{\it " There is some degree of arbitrariness in assigning values to the individual coupling constants
required to calculate radiative decays"}.
In other words, at the present time there are no reference values
for those widths and various calculations put forward the relative importance of mechanisms considered
in each approach. Moreover, given that the $\Lambda$(1405) is 27 MeV below the $K^-p$
threshold, in kaonic atom only the upper tail of that resonance intervenes.
%
%
\begin{table}[ht]
\caption{\footnotesize Radiative decay widths (in keV) of the
$\Lambda(1405)\to {\Lambda \gamma}, ~{\Sigma \gamma}$ decays in different approaches,
and the corresponding ratios
$R=\Gamma _{\Sigma \gamma} / \Gamma _{\Lambda \gamma }$.
}
\begin{tabular} {lrrcccl}
\hline \hline
Approach & $\Gamma _{\Lambda \gamma}$  & $\Gamma _{\Sigma \gamma}$ & & $R$ & Reference \\
\hline
$\chi$QM  & 123 & 212 & & 1.72 & Present work, with $P_{5q}$=45\% \\
$\chi$QM  & 168 & 103 & & 0.61 & Yu {\it et al.}~\cite{Yu:2006sc} \\
Algebric model & 117 & 156 & & 1.33 & Bijker {\it et al.}~\cite{Bijker:2000gq} \\
$U \chi PT$ &  16 &  73 & & 4.56 & Geng {\it et al.}~\cite{Geng:2007hz} \\
  &  65 &  33 & & 0.51  & Geng {\it et al.}~\cite{Geng:2007hz} \\
$U \chi PT$ &  19 &  113 & & 5.95 & Doring {\it et al.}~\cite{Doring:2010fw} \\
            &  83 &   55 & & 0.66  & Doring {\it et al.}~\cite{Doring:2010fw} \\
Bonn CQM       & 912 & 233 & & 0.26 & Van Cauteren {\it et al.}~\cite{VanCauteren:2005sm} \\
NRQM       & 143 &  91 & & 0.64 & Darewych {\it et al.}~\cite{Darewych:1983yw} \\
NRQM       & 154 &  72 & & 0.47 & Kaxiras {\it et al.}~\cite{Kaxiras:1985zv} \\
           & 200 &  72 & & 0.36 & Kaxiras {\it et al.}~\cite{Kaxiras:1985zv} \\
RCQM       & 118 &  46 & & 0.39 & Warns {\it et al.}~\cite{Warns:1990xi} \\
MIT bag    &  60 &  18 & & 0.30 & Kaxiras {\it et al.}~\cite{Kaxiras:1985zv} \\
            &  17 &  3 & & 0.18 & Kaxiras {\it et al.}~\cite{Kaxiras:1985zv} \\
Chiral bag &  75 &  2 & & 0.03  & Umino - Myhrer~\cite{Umino:1992hi} \\
Soliton    &  40 & 17 & & 0.43 & Schat {\it et al.}~\cite{Schat:1994gm} \\
           &  44 &  13 & & 0.30 & Schat {\it et al.}~\cite{Schat:1994gm} \\
 Isobar model & 27 $\pm$ 8 & 10 $\pm$ 4 & & 0.37 $\pm$ 0.18 & Burkhardt - Lowe~\cite{Burkhardt:1991ms} \\
              & 27 $\pm$ 8 & 23 $\pm$ 7 & & 0.85 $\pm$ 0.36 & Burkhardt - Lowe~\cite{Burkhardt:1991ms} \\
\hline \hline
\end{tabular}
\label{others}
\end{table}

Predictions for both channels decay widths (Table \ref{others}) may vary by two orders of magnitude from one approach to another, making any conclusive comparisons pointless in the
absence of data.
Landberger~\cite{Landsberg:1996gb} suggested that the predicted ratio
$R=\Gamma _{\Sigma \gamma} / \Gamma _{\Lambda \gamma }$ might be more
instructive. Inspection of that ratio for different approaches (fourth column in Table \ref{others})
allows us to distinguis three ranges:
$ R  \gtrsim 1.0$ (present work and Refs.~\cite{Geng:2007hz,Bijker:2000gq,Burkhardt:1991ms}),
$0.4 \lesssim  R \lesssim 0.6$ (Refs.~\cite{Geng:2007hz,Yu:2006sc,Darewych:1983yw,Kaxiras:1985zv,Warns:1990xi,Schat:1994gm,Burkhardt:1991ms}),
and $ R \lesssim 0.3$ (Refs.~\cite{VanCauteren:2005sm,Schat:1994gm,Kaxiras:1985zv,Umino:1991dk,Umino:1992hi}).

Our model gives $R$=1.7, almost 29\% larger than that obtained with
algebric model~\cite{Bijker:2000gq}, but about two times smaller
than the ratio given by a very recent coupled channels unitary
chiral perturbation theory ($U \chi PT$)~\cite{Geng:2007hz}. This
latter generates 2 poles corresponding to the nominal
$\Lambda$(1405), resulting in two different radiative decay widths.
The low-energy pole leads to $R$=4.56, with $\Gamma _{\Lambda \gamma
}$=16 keV, compatible with the value extracted within the above
mentioned isobar model~\cite{Burkhardt:1991ms}. However, that model
leads to a ratio compatible, within 1-$\sigma$, with both
$\approx$1.2 and $\approx$0.5, so within the first two ranges. The
results for $ R  \gtrsim 1.0$ lead then to two series with respect
to the width $\Gamma _{\Lambda \gamma }$$\approx$100 keV (present
work and Ref.~\cite{Bijker:2000gq})  and $\approx$20 keV
~\cite{Geng:2007hz,Burkhardt:1991ms}, while $\Gamma _{\Sigma \gamma
}$ varies by two orders of magnitude.

The higher-energy pole in the $U \chi PT$~\cite{Geng:2007hz} comes out in the second range
$0.4 \lesssim  R \lesssim 0.6$. It is worth noticing that various quark model based
approaches~\cite{Darewych:1983yw,Kaxiras:1985zv,Yu:2006sc,Warns:1990xi} predict ratios in the same interval,
and three of them ~\cite{Darewych:1983yw,Kaxiras:1985zv,Warns:1990xi} give close enough predictions for
both $\Gamma _{\Lambda \gamma }$ and $\Gamma _{\Sigma \gamma }$.
It is however known that those approaches fail in describing the $\Lambda$(1405).
In the last part of this section we will come back to the recent chiral quark approach~\cite{Yu:2006sc}.

Moreover, in the MIT bag model~\cite{Kaxiras:1985zv} there are two $J^P={1/2}^-~\Lambda$ states at 1364 MeV
and 1446 MeV, leading to $R$= 0.18 and 0.30, respectively, with decay widths much smaller than those
predicted by quark models, but closer to the
Soliton models~\cite{Schat:1994gm,Kaxiras:1985zv}.
A more advanced chiral approach~\cite{Umino:1991dk,Umino:1992hi}
including gluon exchange mechanism, predicts a larger width for $\Gamma _{\Lambda \gamma }$ (=75 keV),
but that for $\Gamma _{\Sigma \gamma }$ shrinks down to 2 keV. That work reproduces well enough the total
width of $\Lambda$(1520), but underestimates that for $\Lambda$(1405).

Now, we would like to proceed to more detailed comparisons between our results set and that reported
by Yu {\it et al.}~\cite{Yu:2006sc}, also within a chiral quark approach. Here, we need to go back to
Eqs. (\ref{cmc1}) to (\ref{cmc4}). Table \ref{Yu} summarizes the state assignments used in the present
work and those in Ref.~\cite{Yu:2006sc}, showing that in this latter work all resonances have been replaced
by the lowest mass relevant baryon.
The drawback of that approximation on numerical results is presented below.
%
%
\begin{table}[ht!]
\caption{\footnotesize Resonance assignments (see Eqs. (\ref{cmc1}) to (\ref{cmc4})).}
\begin{tabular} {ccc}
\hline \hline
State & Baryon &  Baryon \\
      & Present work &  Ref. \cite{Yu:2006sc} \\
\hline
$\Lambda^{2}_{1}P_{A}$                                                        & $\Lambda^*$(1405)& $\Lambda^*$(1405) \\
$\Lambda^{2}_{8}P_{M}$                                                        & $\Lambda^*$(1670)& $\Lambda^*$(1405) \\
$\Lambda^{4}_{8}P_{M}$                                                        & $\Lambda^*$(1800)& $\Lambda^*$(1405) \\
$\Lambda^{2}_{8}S_{S}$                                                        & $\Lambda$(1116)& $\Lambda$(1116) \\
$\Lambda^{2}_{8}S_{S^\prime}$                                                 & $\Lambda^*$(1600)& $\Lambda$(1116) \\
$\Lambda^{2}_{8}S_{M}$                                                        & $\Lambda^*$(1810)& $\Lambda$(1116) \\
$\Sigma^{2}_{8}S_{S}$                                                         & $\Sigma$(1193) & $\Sigma$(1193)\\
$\Sigma^{2}_{8}S_{S^\prime}$                                                  & $\Sigma^*$(1660) & $\Sigma$(1193)\\
$\Sigma^{2}_{8}S_{M}$                                                         & $\Sigma^{*}$(1770) & $\Sigma$(1193)\\
$N^{2}_{8}S_{S}$                                                              & $N$(938)       & $N$(938) \\
$N^{2}_{8}S_{S^\prime}$                                                       & $N^*$(1440)    & $N$(938) \\
$N^{2}_{8}S_{M}$                                                              & $N^*$(1710)    & $N$(938) \\
\hline \hline
\end{tabular}
\label{Yu}
\end{table}
%
%
The hereafter called hybrid model ($HM$) results are obtained using our code, but state assignments of Yu {\it et al.}~\cite{Yu:2006sc}. The resulting widths are reported
in Table \ref{Yuw}. The width $\Gamma _{\Lambda \gamma }$ increases by about 15\% in going from for pure $3q$ 
configuration to $P_{5q}\lesssim$45\%,
while $\Gamma _{\Sigma \gamma }$ almost doubles, and the ratio $R$ increases rather smootly. 
Although the ratio (0.6) found for
$P_{5q}=$0\% is very close to that obtained by Yu {\it et al.}~\cite{Yu:2006sc}, there are about 25\%
discrepancies among the widths. We will come back to this point.

%
\begin{table}[ht]
\caption{\footnotesize Same as Table \ref{rdw}, but for hybrid model (using our formalism with resonance asignments of Ref. \cite{Yu:2006sc}).}
\begin{tabular} {lccccc}
\hline \hline
& A & B & C & D & E \\
$P_{5q}$ (\%)& 0 & 25 & 45 & 75 & 100 \\
 \hline

$\Gamma_{\Lambda \gamma}$ & 119 & 141 & 134 &  105 & 47  \\

$\Gamma_{\Sigma \gamma}$ &   77 & 158 & 169 & 158 & 102  \\

$R$ &   0.6 & 1.1 & 1.3 & 1.5 & 2.2  \\

\hline \hline
\end{tabular}
\label{Yuw}
\end{table}
%

%
\begin{table}[ht!]
\caption{\footnotesize Numerical results for the helicity amplitude $A^{\Lambda}_{1/2}$ (in
GeV$^{-1/2}$) for electromagnetic transition
$\gamma\Lambda\rightarrow\Lambda(1405)$, with our results (2nd column), those from the hybrid model
($HM$, 3rd column), and from Yu {\it et al.}~\cite{Yu:2006sc} (last column).}
\begin{tabular} {lccc}
\hline \hline
&  total & $HM$ & Ref. \cite{Yu:2006sc}\\
\hline
$\Lambda^{2}_{8}S_{S}~\to~\Lambda^{2}_{1}P_{A}$ & 0.087  & 0.087  & -0.070\\
$\Lambda^{2}_{8}S_{S}~\to~\Lambda^{2}_{8}P_{M}$ & -0.021 & -0.032 & 0.062\\
$\Lambda^{2}_{8}S_{S}~\to~\Lambda^{4}_{8}P_{M}$ & 0.016 & 0.0013& -0.004\\

$\Lambda^{2}_{8}S'_{S}~\to~\Lambda^{2}_{1}P_{A}$ & -0.008 & -0.006 & 0.030\\
$\Lambda^{2}_{8}S'_{S}~\to~\Lambda^{2}_{8}P_{M}$ & -0.025 & -0.012 & -0.035\\
$\Lambda^{2}_{8}S'_{S}~\to~\Lambda^{4}_{8}P_{M}$ & -0.009 & 0.006 & -0.002\\

$\Lambda^{2}_{8}S_{M}~\to~\Lambda^{2}_{1}P_{A}$ & -0.011 & -0.019 & -0.021\\
$\Lambda^{2}_{8}S_{M}~\to~\Lambda^{2}_{8}P_{M}$ & -0.016 & -0.015 &-0.008\\
$\Lambda^{2}_{8}S_{M}~\to~\Lambda^{4}_{8}P_{M}$ &  -0.053 & -0.009 & -0.002\\
\hline \hline
\end{tabular}
\label{compe}
\end{table}
%
%
\begin{table}[ht]
\caption{\footnotesize Results for
the helicity amplitude $A^{\Sigma}_{1/2}$ (in GeV$^{-1/2}$)
of electromagnetic transitions
$\gamma\Sigma\rightarrow\Lambda(1405)$. Columns are as in Table \ref{compe}.}
\begin{tabular} {lccc}
\hline \hline
 & total & $HM$&  Ref. \cite{Yu:2006sc}\\
\hline
$\Sigma^{2}_{8}S_{S}~\to~\Lambda^{2}_{1}P_{A}$ & -0.205 & -0.205 & -0.216\\
$\Sigma^{2}_{8}S_{S}~\to~\Lambda^{2}_{8}P_{M}$ & -0.093 & -0.146 & -0.202\\
$\Sigma^{2}_{8}S_{S}~\to~\Lambda^{4}_{8}P_{M}$ & 0.003 & -0.032 &  0.007\\
$\Sigma^{2}_{8}S'_{S}~\to~\Lambda^{2}_{1}P_{A}$ & 0.011 & 0.018 & 0.196\\
$\Sigma^{2}_{8}S'_{S}~\to~\Lambda^{2}_{8}P_{M}$ & -0.075& -0.014  & 0.109\\
$\Sigma^{2}_{8}S'_{S}~\to~\Lambda^{4}_{8}P_{M}$ & -0.091 & -0.043 &  0.004\\
$\Sigma^{2}_{8}S_{M}~\to~\Lambda^{2}_{1}P_{A}$ & 0.027 & 0.047    & -0.074\\
$\Sigma^{2}_{8}S_{M}~\to~\Lambda^{2}_{8}P_{M}$ & -0.108 & -0.076 & 0.005\\
$\Sigma^{2}_{8}S_{M}~\to~\Lambda^{4}_{8}P_{M}$ & -0.204 & -0.074 & 0.003\\
\hline \hline
\end{tabular}
\label{comps}
\end{table}

In Tables \ref{compe} and \ref{comps} we report our results for helicity amplitudes for each state,
including those obtained using the hybrid model, and compare them with values found in Ref \cite{Yu:2006sc}.
Notice that there is an overall sign difference between our conventions and those used in Ref \cite{Yu:2006sc}.
The first observation is that the state assignments of Ref \cite{Yu:2006sc}, affect almost all the amplitudes
for $\gamma^{*}\Lambda\rightarrow\Lambda(1405)$, bringing them closer to those in Ref \cite{Yu:2006sc}.
Then, the fact that the hybrid model and Ref. \cite{Yu:2006sc} produce different results for the decay width can
be attributed on the one hand to small differences in some of the amplitudes and on the other hand to the input values.

The situation is very different for the $\gamma^{*}\Sigma\rightarrow\Lambda(1405)$ transition (Table \ref{comps}).
Although the $HM$ results show significant deviations from our original values, they also differ
very significantly from values reported in Ref \cite{Yu:2006sc}. The main explanation for that feature might be due
to a sign difference in their expression for $\Phi^{\rho}_{\Sigma^\circ}$ (Eq. (A1) in that reference)
and $|\Sigma^{\circ}\rangle_{\rho}$ in the present manuscript (Eq. (\ref{A5})).
This observation explains, at least partly, the differences between
the values found for $\Gamma _{\Sigma \gamma }$ in the result coming from hybrid model and
those reported in Ref \cite{Yu:2006sc}. Results from this latter work, after correcting the sign,
might allow more conclusive comparisons with our findings.

At this point, and having discussed results compiled in Table \ref{others}, the main firm message is that
decay widths measurements are mandatory in identifying the most reliable approaches. In the meantime,
comparisons among outputs from those works with other observables constitute an alternative way to progress.
Accordingly, in the next Section we concentrate on the strong channels decay.

%
\subsection{Strong decay of $\Lambda(1405)$}
\label{nrsd}
Using the formalism developed in Sec. \ref{fsd} and transition amplitudes reported in
Tables \ref{sigpi} and \ref{kn}, here we present our numerical results.

The transition amplitudes for $\Lambda(1405)\to \Sigma(1194)\pi$
and $\Lambda(1405)\to K^{-}p$ are given in Tables \ref{aps} and \ref{akp}, respectively.

%
\begin{table}[ht]
\caption{\footnotesize Results for the amplitudes of the
$\Lambda(1405) \to \Sigma(1194) \pi$ decay. Amplitudes for $5q \to 5q$ transitions vanish (see Sec. \ref{wf5q}).}
\begin{tabular} {lcccc}
\hline \hline
& $3q \to 3q$ &  N-D & total\\
\hline
$\Lambda^{2}_{1}P_{A}~\to~\Sigma^{2}_{8}S_{S}$    & 0.736    & 0.384    & 1.120 \\
$\Lambda^{2}_{8}P_{M}~\to~\Sigma^{2}_{8}S_{S}$    & 0.287    & -0.229   & 0.058 \\
$\Lambda^{4}_{8}P_{M}~\to~\Sigma^{2}_{8}S_{S}$    & 0.491    & -0.204   & 0.287 \\
$\Lambda^{2}_{1}P_{A}~\to~\Sigma^{2}_{8}S'_{S}$   & -0.722   & 0        & -0.722 \\
$\Lambda^{2}_{8}P_{M}~\to~\Sigma^{2}_{8}S'_{S}$   &  0.001   & 0        & -0.001 \\
$\Lambda^{4}_{8}P_{M}~\to~\Sigma^{2}_{8}S'_{S}$   & -0.228   & 0        & -0.228 \\
$\Lambda^{2}_{1}P_{A}~\to~\Sigma^{2}_{8}S_{M}$    & -0.511   & 0        & -0.511 \\
$\Lambda^{2}_{8}P_{M}~\to~\Sigma^{2}_{8}S_{M}$    & -0.051   & 0        & -0.051 \\
$\Lambda^{4}_{8}P_{M}~\to~\Sigma^{2}_{8}S_{M}$    & -0.016   & 0        & -0.016 \\
\hline \hline
\end{tabular}
\label{aps}
\end{table}
%

%
%
\begin{table}[hb!]
\caption{\footnotesize Results for
the amplitudes of the $\Lambda(1405)\to K^{-}p$ decay.
Amplitudes for $5q \to 5q$ transitions vanish (see Sec. \ref{wf5q}).}
\begin{tabular} {lcccc}
\hline \hline
& $3q \to 3q$ &  $N-D$ & total\\
\hline
$\Lambda^{2}_{1}P_{A}~\to~N^{2}_{8}S_{S}$    & 1.478     & -0.824    & 0.654 \\
$\Lambda^{2}_{8}P_{M}~\to~N^{2}_{8}S_{S}$    & -0.878    & -0.228    & -1.106  \\
$\Lambda^{4}_{8}P_{M}~\to~N^{2}_{8}S_{S}$    & 0         & -0.198    & -0.198 \\

$\Lambda^{2}_{1}P_{A}~\to~N^{2}_{8}S'_{S}$   & 0.745     & 0        & 0.745 \\
$\Lambda^{2}_{8}P_{M}~\to~N^{2}_{8}S'_{S}$   & -0.140    & 0        & -0.140 \\
$\Lambda^{4}_{8}P_{M}~\to~N^{2}_{8}S'_{S}$   & 0         & 0        & 0      \\

$\Lambda^{2}_{1}P_{A}~\to~N^{2}_{8}S_{M}$    & 0.082     & 0        & 0.082      \\
$\Lambda^{2}_{8}P_{M}~\to~N^{2}_{8}S_{M}$    & -0.363    & 0        & -0.363 \\
$\Lambda^{4}_{8}P_{M}~\to~N^{2}_{8}S_{M}$    & -0.194    & 0        &  -0.194 \\
\hline
\end{tabular}
\label{akp}
\end{table}
%

The nondiagonal terms, wherever relevant, play significant roles in both decay channels.
For the $\Lambda(1405) \to \Sigma(1194) \pi$ transition (Table \ref{aps}) the effect turns out
to be constructive for the first transition, $\Lambda^{2}_{1}P_{A}~\to~\Sigma^{2}_{8}S_{S}$,
enhancing its dominant character.
For the two other transitions the destructive combinations of those terms with pure $3q$ transitions lead
to almost vanishing contribution from $\Lambda^{2}_{8}P_{M}~\to~\Sigma^{2}_{8}S_{S}$, suppressed
by a factor of 2, the mgnitude of $\Lambda^{4}_{8}P_{M}~\to~\Sigma^{2}_{8}S_{S}$ transition amplitude.

For the $\Lambda(1405)\to K^{-}p$ decay (Table \ref{akp}), the dominant term in pure $3q$ transition,
$\Lambda^{2}_{1}P_{A} \to N^{2}_{8}S_{S}$, gets reduced by more than 50\% due to the nondiagonal term,
while the magnitude of the second transition, $\Lambda^{2}_{8}P_{M} \to N^{2}_{8}S_{S}$, increases by 20\%.
Finally, the nondiagonal terms attribute a significant role to the
$\Lambda^{4}_{8}P_{M} \to N^{2}_{8}S_{S}$ transition, otherwise vanishing in pure $3q \to 3q$ scheme.

Using those transition amplitudes, we now move to numerical results for decay width and coupling
constants.
In Table \ref{strong}, we give the numerical results with
$P_{5q}=0\%,25\%,45\%,75\%$ and $100\%$ in columns A, B, C, D and E, respectively.
By comparing results in columns $A$ and $B$, we observe very
significant effects arising from the nondigonal terms discussed above.

%
\begin{table}[ht!]
\caption{\footnotesize Results for the $\Sigma\pi$ decay width of $\Lambda(1405)$, and the
$\Lambda(1405)\Sigma\pi$ and $\Lambda(1405) K^{-}p$ couplings.}
\begin{tabular} {lccccc}
\hline \hline
&  A    & B  & C  & D  & E  \\
$P_{5q}$ (\%)                        &  0  & 25 & 45 & 75 & 100  \\
\hline
$\Gamma_{\Sigma\pi}$ (MeV)           & 24  & 47 & 50 & 45 & 23   \\
$f_{\Lambda(1405)\Sigma\pi}/m_{\pi}$ & 3.0 &4.1 &4.3 &4.1 &2.9   \\
$f_{\Lambda(1405)K^{-}p}/m_{K}$      & 11.3& 7.4 &5.4 &1.9 &-4.1 \\
\hline \hline
\end{tabular}
\label{strong}
\end{table}
%

The most striking result is the predicted values for the width of $\Lambda(1405) \to \Sigma\pi$ decay.
While a pure $3q$ constituent quark model underestimates that observable by a factor of 2,
introduction of five-quark components in $\Lambda(1405)$ with $P_{5q} \approx$50\%, leads to excellent
agreement with the value, 50$\pm$2, reported in PDG \cite{Amsler:2008zzb}, and coming from
Ref. \cite{Dalitz:1991sq}. This latter work, published by Dalitz and Deloff in 1991,
is an impulse approximation approach fitting a subset of data from Ref. \cite{Hemingway:1984pz},
and discarding the only other data set \cite{Thomas:1973uh} available at that time.

In Table \ref{strong-s}, we summarize the relevant works on $\Gamma_{\Lambda(1405) \to \Sigma\pi}$.
Recent data obtained at COSY by Zychor {\it et al.} \cite{Zychor:2007gf} give a decay width of
about 60 MeV, and a recent \cite{Esmaili:2009rf} phenomenological analysis leads to 40$\pm$8.
Two other formalisms, based on Bethe-Salpeter coupled-channels \cite{GarciaRecio:2002td}
and chiral quark model \cite{Zhong:2008km}, find values compatible with the findings by
Dalitz and Deloff \cite{Dalitz:1991sq}. Our result is also in line with those reported values.
Width determined within a unitary chiral perturbation theory \cite{Magas:2005vu} suggests a smaller value,
within a double-pole picture of $\Lambda(1405)$. Very recently Akaishi {\it et al.} \cite{Akaishi:2010wt},
using a varational treatment, questioned that picture and advocated a single-pole nature for that resonance.

So, within our work with $P_{5q}$=45\%, the $\Lambda(1405)$ resonance appears to favor a mixed
structure of the three- and five-quark components.

%
\begin{table}[ht!]
\caption{\footnotesize Results for
the $\Sigma\pi$ decay width of $\Lambda(1405)$.}
\begin{tabular} {lcl}
\hline \hline
 Approach    & $\Gamma_{\Lambda(1405)\to (\Sigma \pi)^\circ}$ & Reference \\
\hline
$\chi QM$ & 50 & Present work with $P_{5q}$ 45\% \\
Bethe-Salpeter coupled-channels& 50$\pm$7 & Garcia-Recio {\it et al.} \cite{GarciaRecio:2002td} \\
$U\chi PT$&  38 & Magas {\it et al.} \cite{Magas:2005vu}  \\
$\chi QM$ &  48 & Zhong - Zhao  \cite{Zhong:2008km} \\
coupled-channels potential model&  40$\pm$8 & Esmaili {\it et al.} \cite{Esmaili:2009rf} \\
COSY experiment & $\approx$ 60 & Zychor {\it et al.} \cite{Zychor:2007gf}  \\
K-matrix & 50$\pm$2 & Dalitz - Deloff \cite{Dalitz:1991sq}, PDG \cite{Amsler:2008zzb}\\
\hline \hline
\end{tabular}
\label{strong-s}
\end{table}

Finally, our results for the $\Lambda(1405)\Sigma\pi$ and
$\Lambda(1405)K^{-}p$ couplings (Table \ref{strong},
reported without including isospin factors,
show significantly different dependence on the structure of $\Lambda(1405)$, namely, in going
from a pure $3q$ configuration to an admixture of the three- and five-quark components,
the coupling ${f_{\Lambda(1405)\Sigma\pi}}$ gets increased by roughly 30\%,
while ${f_{\Lambda(1405)K^-P}}$ decreases by about 40\%.

\section{Summary and conclusions}
\label{fn}

Within an extended chiral constituent quark model, we investigated the three- and five-quark
structure of the $S_{01}$ resonance $\Lambda(1405)$.
The wave functions for this resonance and the octet baryons in our approach were reported explicitly.
We derived the electro-excitation helicity amplitudes for
$\gamma ^* \Lambda(1116) \to \Lambda(1405)$, $\gamma ^* \Sigma ^\circ (1194) \to \Lambda(1405)$
processes, as well as transition amplitudes for the $\Lambda(1405)\to\Sigma(1194)\pi$, $ K^{-}p$ decays.
Using those amplitudes, we gave expressions for the electromagnetic
and strong decays widths, namely,
$\Gamma_{\Lambda(1405) \to Y \gamma}$, with $Y \equiv \Lambda(1116),~\Sigma(1194)$ and
$\Gamma_{\Lambda(1405)\to {(\Sigma \pi)}^\circ}$,
with $(\Sigma \pi)^\circ \equiv \Sigma^\circ \pi^\circ,~\Sigma^+ \pi^-,~\Sigma^- \pi^+$.

The numerical values computed using those expressions were presented and
the dependence of various decay widths on the percentage of the five-quark components were investigated
and compared with other sources. For the photo-excitation helicity amplitudes $A^{\Lambda} _{1/2}$,
we found good agreements with the only set of published results  by Yu {\it et al.} \cite{Yu:2006sc},
using their approximations.
For the $A^{\Sigma} _{1/2}$, a seemingly sign problem in that paper did not allow us to proceed to
meaningful comparisons.
We also examined the situation with respect to the decay widths
$\Gamma _{\Lambda(1405)\to\Lambda \gamma}$,
$\Gamma _{\Lambda(1405)\to\Sigma \gamma}$ and their ratio.
We argued that large discrepencies among a dozen 
works \cite{Darewych:1983yw,Kaxiras:1985zv,VanCauteren:2005sm,Yu:2006sc,Schat:1994gm,Burkhardt:1991ms,Bijker:2000gq,Geng:2007hz,Warns:1990xi,Umino:1991dk,Umino:1992hi}
devoted to that topic render it impossible
to make any conclusive comparisons.
Then, among the quanties investigated here, the only firm ground is offered by the experimental
results for the $\Lambda(1405)\to\Sigma(1194)\pi$ decay width ($\Gamma_{(\Sigma\pi)^\circ}$).
Our formalism, embodying about 45\% of five-quark components in the $\Lambda(1405)$
resonance and $20\%$ in the octet baryons, allows reproducing $\Gamma_{(\Sigma\pi)^\circ}=50\pm2$
reported in PDG and endorsed OUR other findings, especially with respect to the electromagnetic decay widths.

Our work hence favors a mixed
structure of the three- and five-quark components in the $\Lambda(1405)$ resonance,
with $[31]_{XFS}[4]_{X}[211]_{F}[22]_{S}$ scheme for the orbital-flavor-spin
configuration of the four-quark subsystem.
This configuration allows
the presence of the $u\bar{u}$, $d\bar{d}$ and $s\bar{s}$ components in
$\Lambda(1405)$, while as shown by An {\it et al.} \cite{An:2008xk,An:2008tz},
that configuration rules out the $u\bar{u}$
and $d\bar{d}$ components in $N(1535)$. Moreover, the probability of
the five-quark components in $N(1535)$ turns out to be in the same range as that
of $\Lambda(1405)$, making the $N(1535)$ heavier than
$\Lambda(1405)$.
In consequence, with respect to the five-quark components in baryons, our results complementing
those published on the Roper \cite{Li:2006nm,JuliaDiaz:2006av} and the
first $S_{11}$ resonances \cite{An:2008xk,An:2009uv}, allows us to put forward an explanation
for the mass ordering of the $N(1440)$, $\Lambda(1405)$,
and $N(1535)$ resonances.
Those issues have also been investigated in lattice QCD approaches \cite{Lee:2002gn,Mathur:2003zf},
an effective linear realization chiral $SU_L(2)$ x $SU_R(2)$ and $U_A(1)$ symmetric Lagrangian \cite{Dmitrasinovic:2009vy}, and concisely reviewed in \cite{Zou:2009zz}.

Finally, we wish to underline the importance of the mixing mechanism resulting from the present study.
The presence of three- and five-qurak components in $\Lambda(1405)$ leads to nondiagonal terms
arising from transitions among those components ($qqq \leftrightarrow qqqq_i{\bar q_i}$).
In the case of photo-excitation helicity amplitudes, we find larger effects due to those transitions than
contributions from five-quark components. For the strong channels, not getting any contributions from those
pure five-quark components, the nondiagonal terms turn out again to be crucial, increasing by about a factor
of 2 the width for the $\Lambda(1405)\to \Sigma(1194)\pi$ decay and bringing it into agreement with the data.
 Comparable effects due to the mixing mechanism have also been reported
for the electromagnetic transition
$\gamma^{*}N\rightarrow N(1535)$ \cite{An:2008xk,An:2009uv}, and the
radiative and strong decays of the Roper resonance \cite{Li:2006nm,JuliaDiaz:2006av}.
This may reveal a new mechanism for the decay properties of baryons, i.e.
$q\bar{q} \to \gamma^*,~\pi,~K$ transitions have significant
contributions to the baryon resonance decays.


\begin{appendix}

\section{Wave functions for the three quark components}
\label{tq}
\subsection{Flavor wave functions}
\label{tqf}
The flavor wave functions for the baryons considered in this paper are as follows:
\begin{eqnarray}
|\Lambda\rangle_{a}&=&\frac{1}{\sqrt{6}}\{|uds\rangle+|dsu\rangle+|sud\rangle
-|usd\rangle-|dus\rangle-|sdu\rangle\}\,,
\label{A1}
\\
|\Lambda\rangle_{\rho}&=&\frac{1}{2\sqrt{3}}\{|usd\rangle-|dsu\rangle-|sud\rangle
+|sdu\rangle+2|uds\rangle-2|dus\rangle\}\,,
\label{A2}\\
|\Sigma^{\circ}\rangle_{\lambda}&=&-\frac{1}{2\sqrt{3}}\{|usd\rangle+|dsu\rangle
+|sud\rangle+|sdu\rangle-2|uds\rangle-2|dus\rangle\}\,,
\label{A3}\\
|\Lambda\rangle_{\lambda}&=&\frac{1}{2}\{|usd\rangle+|sud\rangle-|sdu\rangle
-|dsu\rangle\}\,,
\label{A4}\\
|\Sigma^{\circ}\rangle_{\rho}&=&\frac{1}{2}\{|usd\rangle+|dsu\rangle-
|sud\rangle-|sdu\rangle\}\,,
\label{A5}\\
|p\rangle_{\lambda}&=&\frac{1}{\sqrt{6}}\{2|uud\rangle-|duu\rangle-|udu\rangle\}\,,\\
|p\rangle_{\rho}&=&\frac{1}{\sqrt{2}}\{|udu\rangle-|duu\rangle\}.
\end{eqnarray}

\subsection{Spin wave functions}
\label{tqs}
The spin-orbital coupled wave function read
\begin{eqnarray}
X_{a}&=&-|\frac{1}{2},\frac{1}{2}\rangle_{\lambda}(\rho,0)+\sqrt{2}|\frac{1}{2},
-\frac{1}{2}\rangle_{\lambda}(\rho,+1)+|\frac{1}{2},\frac{1}{2}\rangle_{\rho}(\lambda,0)
-\sqrt{2}|\frac{1}{2},-\frac{1}{2}\rangle_{\rho}(\lambda,+1)\,,\\
X_{\lambda}&=&-|\frac{1}{2},\frac{1}{2}\rangle_{\lambda}(\lambda,0)+\sqrt{2}|\frac{1}{2},
-\frac{1}{2}\rangle_{\lambda}(\lambda,+1)+|\frac{1}{2},\frac{1}{2}\rangle_{\rho}(\rho,0)-
\sqrt{2}|\frac{1}{2},-\frac{1}{2}\rangle_{\rho}(\rho,+1)\,,\\
X_{\rho}&=&|\frac{1}{2},\frac{1}{2}\rangle_{\rho}(\lambda,0)-\sqrt{2}|\frac{1}{2},-\frac{1}{2}
\rangle_{\rho}(\lambda,+1)+|\frac{1}{2},\frac{1}{2}\rangle_{\lambda}(\rho,0)-\sqrt{2}|\frac{1}{2},
-\frac{1}{2}\rangle_{\lambda}(\rho,+1)\,,\\
X_{\lambda}^{'}&=&\sqrt{3}|\frac{3}{2},\frac{3}{2}\rangle(\lambda,-1)-\sqrt{2}|\frac{3}{2},\frac{1}{2}\rangle(\lambda,0)
+|\frac{3}{2},-\frac{1}{2}\rangle(\lambda,1)\,,\\
X_{\rho}^{'}&=&\sqrt{3}|\frac{3}{2},\frac{3}{2}\rangle(\rho,-1)-\sqrt{2}|\frac{3}{2},\frac{1}{2}\rangle(\rho,0)
+|\frac{3}{2},-\frac{1}{2}\rangle(\rho,1),
\end{eqnarray}
with $(\lambda,0)=q_{\lambda,z}$,
$(\lambda,+1)=-\frac{1}{\sqrt{2}}(q_{\lambda,x}+iq_{\lambda,y)}$,
$(\rho,0)=q_{\rho,z}$ and
$(\rho,+1)=-\frac{1}{\sqrt{2}}(q_{\rho,x}+iq_{\rho,y})$.

The spin wave functions are
\begin{eqnarray}
|\frac{1}{2},\frac{1}{2}\rangle_{\rho}&=&\frac{1}{\sqrt{2}}\{|\uparrow\downarrow\uparrow\rangle
-|\downarrow\uparrow\uparrow\rangle\},|\frac{1}{2},\frac{1}{2}\rangle_{\lambda}=\frac{1}{\sqrt{6}}
\{2|\uparrow\uparrow\downarrow\rangle-|\uparrow\downarrow\uparrow\rangle-|\downarrow\uparrow\uparrow\rangle\}\,,\\
|\frac{1}{2},-\frac{1}{2}\rangle_{\rho}&=&\frac{1}{\sqrt{2}}\{|\uparrow\downarrow\downarrow\rangle
-|\downarrow\uparrow\downarrow\rangle\},|\frac{1}{2},\frac{1}{2}\rangle_{\lambda}=-\frac{1}{\sqrt{6}}
\{2|\downarrow\uparrow\uparrow\rangle-|\downarrow\uparrow\downarrow\rangle-|\uparrow\downarrow\downarrow\rangle\}\,,\\
|\frac{3}{2},\frac{3}{2}\rangle&=&|\uparrow\uparrow\uparrow\rangle,|\frac{3}{2},\frac{1}{2}\rangle=\frac{1}
{\sqrt{3}}\{|\uparrow\uparrow\downarrow\rangle+|\uparrow\downarrow\uparrow\rangle+|\downarrow\uparrow\uparrow\rangle\}\,,\\
|\frac{3}{2},-\frac{1}{2}\rangle&=&\frac{1}{\sqrt{3}}\{|\uparrow\downarrow\downarrow\rangle+|\downarrow\uparrow\downarrow\rangle
+|\downarrow\downarrow\uparrow\rangle\}.
\end{eqnarray}

\subsection{Orbital wave functions}
\label{tqo}
Here we employ the harmonic oscillator wave functions
\begin{eqnarray}
\Phi_{\Lambda^{*}}(\vec{q}_{\lambda},\vec{q}_{\rho})&=&\frac{\sqrt{2}}{\pi^{3/2}\omega_{3}^{4}}
exp\{-\frac{q_{\lambda}^{2}+q_{\rho}^{2}}{2\omega_{3}^{2}}\}\,,\\
\Phi_{000}(\vec{q}_{\lambda},\vec{q}_{\rho})&=&\frac{1}{(\pi\omega_{3}^{2})^{3/2}}
exp\{-\frac{q_{\lambda}^{2}+q_{\rho}^{2}}{2\omega_{3}^{2}}\}\,,\\
\Phi_{200}^{s}(\vec{q}_{\lambda},\vec{q}_{\rho})&=&\frac{1}{\sqrt{3}(\pi\omega_{3}^{2})^{3/2}}(3-\frac{q_{\lambda}^{2}
+q_{\rho}^{2}}{\omega_{3}^{2}})exp\{-\frac{q_{\lambda}^{2}+q_{\rho}^{2}}{2\omega_{3}^{2}}\}\,,\\
\Phi_{200}^{\rho}(\vec{q}_{\lambda},\vec{q}_{\rho})&=&\frac{2}{\sqrt{3}\pi^{3/2}\omega_{3}^{5}}(\vec{q}_{\rho}\cdot
\vec{q}_{\lambda})exp\{-\frac{q_{\lambda}^{2}+q_{\rho}^{2}}{2\omega_{3}^{2}}\}\,,\\
\Phi_{200}^{\lambda}(\vec{q}_{\lambda},\vec{q}_{\rho})&=&\frac{1}{\sqrt{3}\pi^{3/2}\omega_{3}^{5}}(q_{\rho}^{2}
-q_{\lambda}^{2})exp\{-\frac{q_{\lambda}^{2}+q_{\rho}^{2}}{2\omega_{3}^{2}}\}\
\end{eqnarray}

\section{Wave functions for the five quark components}
\label{fq}

\subsection{Flavor and spin couplings}
\label{fqfs} The decomposition of the flavor-spin configuration
$[31]_{FS}[211]_{F}[22]_{S}$ is \cite{chen}
\begin{eqnarray}
|[31]_{FS}\rangle_{1}&=&\frac{1}{\sqrt{2}}\{|[211]\rangle_{F1}|[22]\rangle_{S1}+
|[211]\rangle_{F2}|[22]\rangle_{S2}\}\,,\\
|[31]_{FS}\rangle_{2}&=&\frac{1}{2}\{-\sqrt{2}|[211]\rangle_{F3}|[22]\rangle_{S2}+
|[211]\rangle_{F2}|[22]\rangle_{S2}-|[211]\rangle_{F1}|[22]\rangle_{S1}\}\,,\\
|[31]_{FS}\rangle_{3}&=&\frac{1}{2}\{[211]\rangle_{F1}|[22]\rangle_{S2}+|[211]\rangle_{F2}|[22]\rangle_{S1}
+\sqrt{2}|[211]\rangle_{F3}|[22]\rangle_{S1}\}\,,
\end{eqnarray}
and that for $[4]_{FS}[22]_{F}[22]_{S}$
\begin{equation}
|[4]_{FS}\rangle=\frac{1}{\sqrt{2}}\{|[22]_{F1}\rangle|[22]_{S1}\rangle+|[22]_{F2}\rangle|[22]_{S2}\rangle\}
\end{equation}

\subsection{Flavor wave functions}
\label{fqf}
The flavor wave functions for $[22]_{F}$ in the $uuds\bar{s}$ component
\begin{eqnarray}
|[22]_{F1}\rangle&=&\frac{1}{\sqrt{24}}\{2|uuds\rangle+2|uusd\rangle+2|dsuu\rangle
+2|sduu\rangle-|duus\rangle-|udus\rangle-|sudu\rangle\nonumber\\
&&-|usdu\rangle-|suud\rangle-|dusu\rangle-|usud\rangle-|udsu\rangle\}\,,\\
|[22]_{F2}&=&\frac{1}{\sqrt{8}}\{|udus\rangle+|sudu\rangle+|dusu\rangle+|usud\rangle
-|duus\rangle-|usdu\rangle-|suud\rangle\nonumber\\
&&-|udsu\rangle\}\,.
\end{eqnarray}

The flavor wave functions for $[211]_{F}$ in the $uuds\bar{s}$ component
\begin{eqnarray}
|[211]\rangle_{F1}&=&\frac{1}{4}\{2|uuds\rangle-2|uusd\rangle-|duus\rangle-|udus\rangle
-|sudu\rangle-|usdu\rangle+|suud\rangle\,\nonumber\\
&&+|dusu\rangle+|usud\rangle+|udsu\rangle\}\,,\\
|[211]\rangle_{F2}&=&\frac{1}{\sqrt{48}}\{3|udus\rangle
-3|duus\rangle+3|suud\rangle-3|usud\rangle
+2|dsuu\rangle-2|sduu\rangle\,\nonumber\\
&&-|sudu\rangle+|usdu\rangle+|dusu\rangle-|udsu\rangle\}\,,\\
|[211]\rangle_{F3}&=&\frac{1}{\sqrt{6}}\{|sudu\rangle+|udsu\rangle
+|dsuu\rangle-|usdu\rangle-|dusu\rangle-|sduu\rangle\}\,.
\end{eqnarray}
All of the other flavor wave functions which are used in this paper are
obtained by applying the lowering operator in the $SU(3)$ flavor space to the above functions.

\subsection{Spin wave functions}
\label{fqs}
Expressions for the spin wave functions $[22]_{S}$ are
\begin{eqnarray}
|[22]\rangle_{S1}&=&\frac{1}{\sqrt{12}}\{2
|\uparrow\uparrow\downarrow\downarrow\rangle
+2|\downarrow\downarrow\uparrow\uparrow\rangle
-|\downarrow\uparrow\uparrow\downarrow\rangle
-|\uparrow\downarrow\uparrow\downarrow\rangle
-|\downarrow\uparrow\downarrow\uparrow\rangle
-|\uparrow\downarrow\downarrow\uparrow\rangle\}\,,\\
|[22]\rangle_{S2}&=&\frac{1}{2}\{
|\uparrow\downarrow\uparrow\downarrow\rangle
+|\downarrow\uparrow\downarrow\uparrow\rangle
-|\downarrow\uparrow\uparrow\downarrow\rangle
-|\uparrow\downarrow\downarrow\uparrow\rangle\}\,.
\end{eqnarray}

\subsection{Orbital wave functions}
\label{fqo}
The orbital wave function of the five-quark components in $\Lambda(1405)$ reads
\begin{eqnarray}
[4]_{X}\Psi(\vec \kappa_i)&=&\frac{1}{\pi^{3}\omega_{5}^{6}}
\exp\{-\frac{\sum_{i}\kappa_{i}^{2}}{2\omega_{5}^{2}}\}.
\end{eqnarray}
The color-orbital coupled wave function for the five-quark
components in the octet baryons is
\begin{eqnarray}
\psi_{C}(\{\vec{\kappa}_i\})&=&\frac{1}{\sqrt{3}}\{[211]_{C1}
\varphi_{01m}(\vec{\kappa}_{1})\varphi_{000}(\vec{\kappa}_{2})
\varphi_{000}(\vec{\kappa}_{3})
-[211]_{C2}\varphi_{000}(\vec{\kappa}_{1})\varphi_{01m}(\vec{\kappa}_{2})
\varphi_{000}(\vec{\kappa}_{3})\,\nonumber\\
&&+[211]_{C3}\varphi_{000}(\vec{\kappa}_{1})\varphi_{000}(\vec{\kappa}_{2})
\varphi_{01m}(\vec{\kappa}_{3})\}\varphi_{000}(\vec{\kappa}_{4})\,
.\ \label{CX}
\end{eqnarray}
Here $[211]_{Ci}$ denote the three color configurations,
$\varphi_{0lm}(\vec{\kappa}_{i})$ the harmonic orbital wave
function with the quantum number $nlm$ and the oscillator frequency
$\omega_{5}$. Notice that the
$\vec{\kappa}_{i}(i=1,2,3)$ generate the 3 configurations of
$[31]_{X}$ in Eqs. (\ref{31xa})-(\ref{31x}).
\end{appendix}

%
%
\end{document}